\documentclass[preprint2]{aastex}
\usepackage{subfigure}
\usepackage{wrapfig}
\usepackage{rotating}
\usepackage{pdflscape}
\begin{document}

\title{Ages of LMC Star Clusters using ASAD$_2$}

\author{Randa S. Asa'd}
\affil{American University of Sharjah, Physics Department, P.O.Box 26666, Sharjah, UAE}
\email{raasad@aus.edu}

\author{Alexandre Vazdekis}
\affil{Instituto de Astrof'sica de Canarias (IAC), E-38200 La Laguna, Tenerife, Spain; Departamento de Astrof'sica, Universidad de La Laguna, E-38205 Tenerife, Spain}

\author{Sami Zeinelabdin} 
\affil{American University of Sharjah, P. O. Box 26666, Sharjah, UAE}

 \begin{abstract}

We use ASAD$_2$, the new version of ASAD (Analyzer of Spectra for Age Determination), to obtain the age and reddening of 27 LMC clusters from full fitting of integrated spectra using different statistical methods ($\chi^{2}$ and K-S test) and a set of stellar population models including GALAXEV and MILES. We show that our results are in good agreement with the CMD ages for both models, and that metallicity does not affect the age determination for the full spectrum fitting method regardless of the model used for ages with log (age/year) $<$ 9. We discuss the results obtained by the two statistical results for both GALAXEV and MILES versus three factors: age, S/N and resolution (FWHM). \\
The predicted reddening values when using the $\chi^{2}$ minimization method are within the range found in the literature for resolved clusters (i.e: $<$ 0.35), however the K-S test can predict E(B$-$V) higher values. The sharp spectrum transition originated at ages around the supergiants contribution, at either side of the AGB peak around log (age/year) 9.0 and log (age/year) 7.8  are limiting our ability to provide values in agreement with the CMD estimates and as a result the reddening determination is not accurate.  We provide the detailed results of four clusters spanning a wide range of ages. ASAD$_2$ is a user-friendly program available for download on the Web and can be immediately used at $http://randaasad.wordpress.com/asad-package/$.
\\
\\
\end{abstract}  

\section{Introduction}

Accurate ages of star clusters provide critical information about the formation history of the host galaxy and particularly its assembly timescales. Our goal in this work is to present the results of, and, offer a user-friendly program which can provide the parameters of the stellar clusters automatically from their integrated spectra. Such a program can be used with large surveys in which the stellar clusters' integrated spectra are obtained, so that important parameters (age and reddening) can be quickly extracted in order to obtain scientific information about the host galaxy. \\
The Large Magellanic Cloud is close enough so that its stellar clusters can be resolved to derive accurate ages, yet far enough to obtain the integrated spectra of these clusters. This makes the LMC stellar clusters ideal for testing the integrated spectra methods of obtaining the ages.\\
Although there are different ways to derive the age, we use the method of the full integrated spectrum fitting. This way we exploit the full information contained in the integrated cluster light, which is the only way to study stellar cluster systems in distant galaxies. Despite the well known age-metallicity degeneracy, \citet {Bica86a} and \citet {Benitez-Llambay12} have shown that metallicity does not play a significant role in the optical range when applying spectral aging methods, hence we apply the method of \citet {AA02, Palma08} of solving for age and reddening as most of our clusters are young (log (age/year) $<$ 9). In \citet{Asad14}, we introduced the Analyzer of Spectra for Age Determination (ASAD) package, that can solve for age and reddening of stellar clusters simultaneously assuming constant metallicity. This has been performed by a $\chi^{2}$ minimization between the observed optical integrated spectra of the clusters and the synthetic model spectra to find the best match.
In this work we introduce ASAD$_2$, the updated version of ASAD, with enhanced features. We use a fixed LMC metallicity Z=0.008 in this work.
In section 2 we briefly describe the data. We summarize the features of ASAD presented in \citet{Asad14} and introduce the new statistical method of ASAD$_2$ in section 3. The new version of our program provides a more extensive set of model libraries for matching, including GALAXEV models (discussed in Section 4.1) and MILES models (discussed in Section 4.2) followed by analysis of error estimates. Reddening predictions are discussed in section 5. In section 6 we discuss the results obtained for four of our clusters. A summary is given in Section 7.

\section{The Data}

The data set used in this work is the one presented in \citet{Asad13}. Twenty LMC clusters were obtained in two observing runs in 2011 with the RC spectrograph on the 4 m Blanco telescope and with the Goodman spectrograph on the SOAR. We obtained integrated spectra by scanning the cluster with the slit starting on the southern edge, with the slit aligned eastwest. To expand our sample, we used seven additional LMC stellar clusters from the literature: Four clusters from \citet{Santos06} and three clusters from \citet{Palma08}. These spectra were kindly provided by the authors. 
Table \ref{Targets Observed} shows the targets observed with a summary of the literature age and reddening.\\ 

 \begin{deluxetable}{|c|c|c|c|c|c|c|c|c|}
\tabletypesize{\scriptsize}
\tablecaption{Targets Observed \label{Targets Observed}}
\tablewidth{0pt}
\tablehead{
\colhead{Name} & \colhead{ Run/Source} &  \colhead{Resolution (\AA)} &  \colhead{S$/$N} &  \colhead{Age$^1$} &  \colhead{Reference}  &  \colhead{ E(B-V)$^2$} &  \colhead{Reference}
}
\startdata
  
NGC1711 & Blanco2011  & 14 & 118 &  7.40  & \cite {Elson+1991} &  0.16 &   \citet {persson_photometric_1983} \\
NGC1856 & Blanco2011  & 14 & 67 &  7.90 & \citet {Hodge(1984)} &  0.21 &   \citet {kerber_physical_2007} \\  
NGC1903 & Blanco2011  & 14 & 28 &  7.85  & \citet {Vallenari98} &  0.16 & \citet {Vallenari98} \\
NGC1984 & Blanco2011  & 14 & 54 &  6.85  & \citet {Hodge(1983)} &  0.14 &    \citet {meurer_ultraviolet_1990} \\
NGC2011 & Blanco2011  & 14 & 46 &  6.78  & \citet {Hodge(1983)} &  0.08 &    \citet {meurer_ultraviolet_1990} \\
NGC2156 & Blanco2011  & 14 & 49 &  7.78  & \citet {Hodge(1983)} &  0.1 &  \citet {persson_photometric_1983} \\
NGC2157 & Blanco2011  & 14 & 79 &  7.60  & \citet {Elson+1991} &  0.10 &  \citet {persson_photometric_1983} \\
NGC2164 & Blanco2011  & 14 & 98 &  7.70  & \citet {Hodge(1983)} &  0.1 &   \citet {persson_photometric_1983} \\
NGC1651 & SOAR2011  & 3.6 & 4 &  9.30  &  \citet {Mould+(1986)} &  0.09  & \citet {Mould+(1986)} \\
NGC1850 & SOAR2011  & 3.6 & 22 & 7.60  &  \citet {Hodge(1983)} &  0.18  &   \citet {Alcaino87} \\
NGC1863 & SOAR2011  & 3.6 & 21 &  7.76  & \citet {Alcaino87} & 0.2  &  \citet {Alcaino87} \\
NGC1983 & SOAR2011  & 3.6 & 16 & 6.90  & \citet {Hodge(1983)}  &  0.09  &    \citet {meurer_ultraviolet_1990} \\
NGC1994 & SOAR2011  & 3.6 & 49 & 6.86  & \citet {Hodge(1983)} &  0.14  &   \citet {meurer_ultraviolet_1990} \\
NGC2002 & SOAR2011  & 3.6 & 18 & 7.20  & \citet {Elson+1991} &  0.12 &  \citet {persson_photometric_1983} \\
NGC2031 & SOAR2011  & 3.6 & 9 &  8.20  & \citet {Dirsch+(2000)} & 0.09  & \citet {Dirsch+(2000)} \\
NGC2065 & SOAR2011  & 3.6 & 33  &  7.85 & \citet {Hodge(1983)} & 0.18  &   \citet {persson_photometric_1983} \\
NGC2155 & SOAR2011  & 3.6 & 10  &  9.40  & \citet {Elson+1988} &  0.02  &  \citet {kerber_physical_2007}\\
NGC2173 & SOAR2011  & 3.6 & 7 & 9.32  &  \citet {Mould+(1986)} &  0.14  & \citet {Mould+(1986)} \\
NGC2213 & SOAR2011  & 3.6 & 7 &  8.95  & \citet {Dacosta85} &  0.09  & \citet {Dacosta85}\\
NGC2249 & SOAR2011  & 3.6 & 7 &  8.82  & \citet{Elson+1988} &  0.01  & \citet {kerber_physical_2007}\\ 
NGC1839  &  \citet {Santos06}  & 14  &  -  &  7.52  & \citet {Alcaino87} &  0.27  &  \citet {Alcaino87} \\
NGC1870  &  \citet {Santos06}  & 14  &  -  &  7.86  & \citet {Alcaino87} &  0.14  &  \citet {Alcaino87} \\
NGC1894  &  \citet {Santos06}  & 14  &  -  &  7.74  & \citet {Dieball00} &  0.1 &  \citet {Dieball00} \\
SL237  &  \citet {Santos06}  & 14 &  -  &  7.43  & \citet {Alcaino87} &  0.17  &  \citet {Alcaino87} \\   
NGC2136  &  \citet {Palma08}  & 17  &  -  &  7.60  & \citet {Hodge(1983)} &  0.10  &  \citet {persson_photometric_1983} \\
NGC2172  &  \citet {Palma08}  & 17  &  -  &  7.78  & \citet {Hodge(1983)} &  0.1  &  \citet {persson_photometric_1983} \\ 
SL234  &  \citet {Palma08}  & 17  &  -  &  7.68  & \citet {Alcaino87} &  0.15  &   \citet {Alcaino87} \\ 
 
\enddata 
\tablenotetext{1}{These are the CMD ages obtained from the literature. The unit is log(age/yr)}
\tablenotetext{2}{These are the E(B-V) obtained from the literature.}
\end{deluxetable}

\section{ASAD Full Spectrum Fitting Tool}

\begin{figure}
\includegraphics[angle=0,scale=0.2]{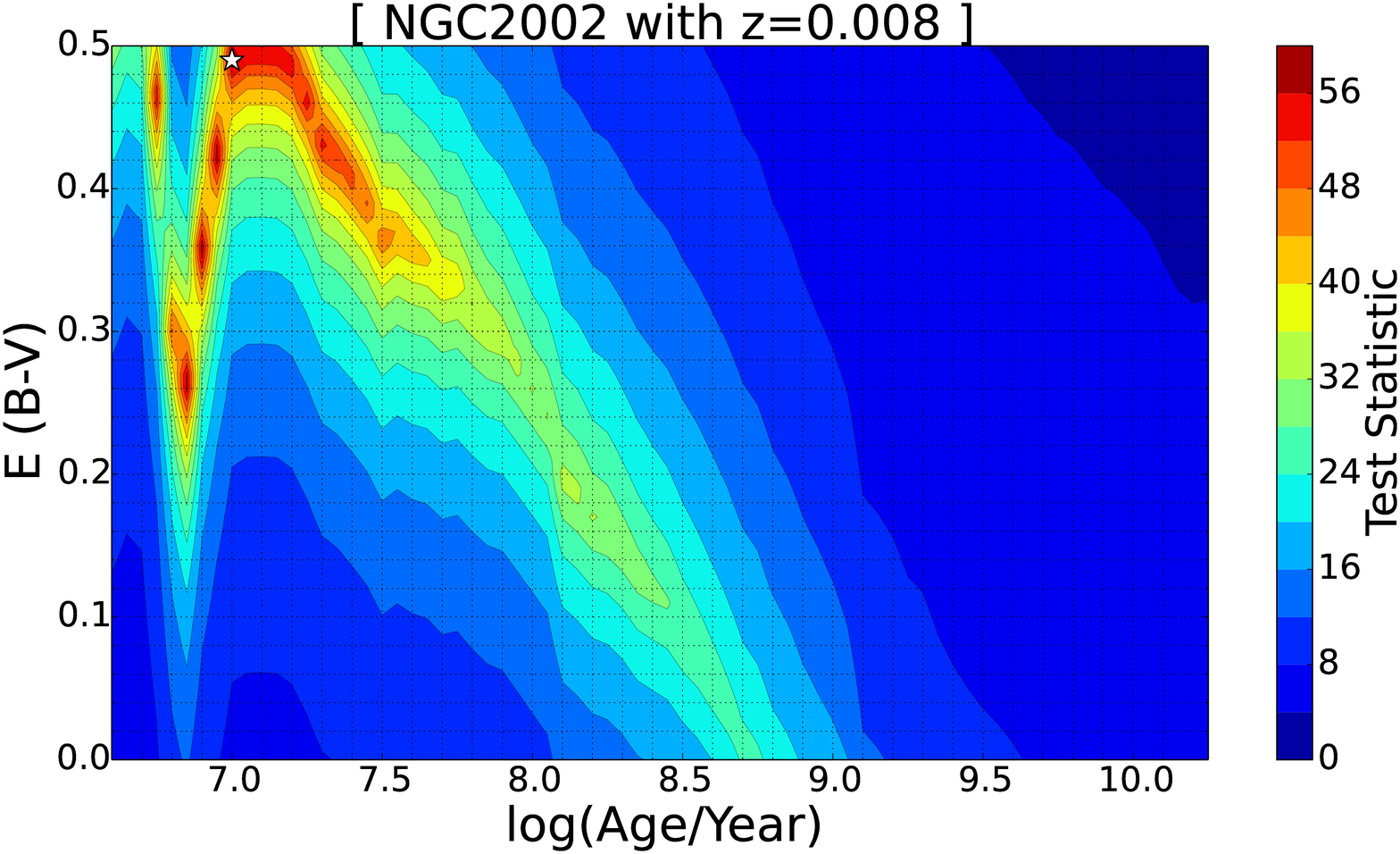}
\caption{The surface plot of NGC2002 predicted by \citet {Delgado05} model with the K-S test. The dark red regions represent the best match. Four solutions for the age/reddening combination are possible. The possible solutions lay in a narrow region of log (Age/year) that is between 6.7 and 7, but a wide region of reddening extending from 0.26 up to 0.49.}
\label{surfaceNGC2002}
\end{figure}

\begin{figure}
\includegraphics[angle=0,scale=0.2]{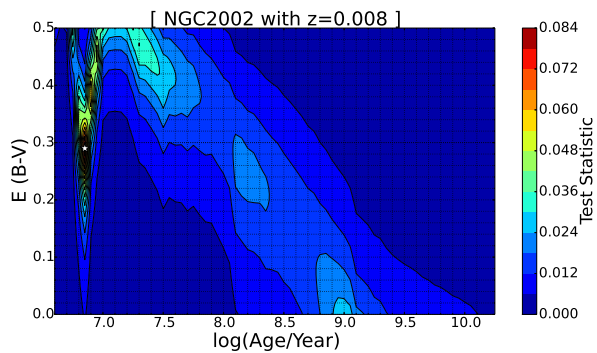}
\caption{The surface plot of NGC2002 using \citet {Delgado05} model with the $\chi^{2}$ minimization method. Only one solution for the age/reddening combination is strongly preferred}
\label{surface_NGC2002.dat_smoothed_SSPPadova.z008.eps}
\end{figure}

\begin{figure}
\includegraphics[angle=0,scale=0.7]{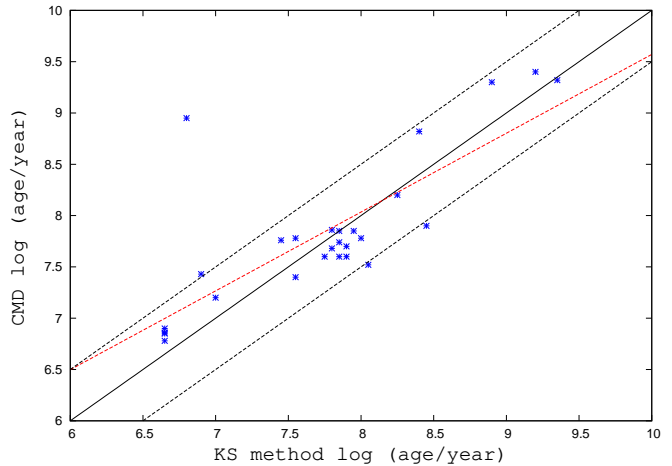}
\caption{The correlation between the ages obtained using the K-S method and the CMD ages. The correlation coefficient is 0.78. The red dashed line is the fit line. See the text for a discussion about the outlier. The dashed lines represent the upper and lower limit of the range of ages within log (age/year) 0.5.}
\label{KS_ages}
\end{figure}

In its first version, ASAD \citep{Asad14} outputs the age and reddening of stellar clusters of known metallicity from their integrated spectra. It performs a $\chi^{2}$ minimization by comparing the observed integrated spectra to the spectral models of \citet {Delgado05}. In this section we will use the same spectral models of \citet {Delgado05} but investigate the method used by \citet{Burke10} to measure the goodness of fit between observed and model spectra, namely the Kolmogorov$-$Smirnov (K-S) test. This test selects the maximum of the absolute value of the difference between the cumulative observed spectrum and the cumulative model spectrum each normalized to unity over the range of wavelengths included in the fit\footnote {ASAD$_2$ allows the user to choose the statistical method preferred ($\chi^{2}$ minimization method or the K-S test method)}. We used the same input parameters as the ones in \citet{Asad13}, a wavelength range of 3626 $-$ 6230 $\rm\AA$, and a step size of 3$\rm\AA$ normalized at 5870$\rm\AA$. The \citet {Cardelli89} extinction law was used with reddening values between 0.00 and 0.50 in steps of 0.01. 
Column 2 in Tables \ref{T2} and \ref{T3} show the results for the best age and reddening value obtained. Column 3 lists the percentage error. It is noticed that although no values of E(B$-$V) higher than 0.35 were found in the literature for the LMC clusters, the K-S test predicts E(B$-$V) values as high as 0.49. 
An investigation of the surface plot of NGC2002, the cluster with the highest model E(B$-$V), is shown in Figure \ref{surfaceNGC2002}. It shows that for NGC2002 many solutions for the age/reddening combination are possible (i.e. dark red regions) based on the K-S test. The possible solutions lay in a narrow region of log (age/year) between 6.7 and 7, with a wide region in reddening extending from 0.26 up to 0.49. Note that there is a significant decrease of the reddening estimate below log (age/year) of 7. This is likely because it corresponds to the peak of the Red Supergiants contributions, which redden the resulting stellar populations spectra. 
When the reddening limit allowed in ASAD$_2$ is expanded to 0.8, the predicted E(B$-$V) gets as high as 0.61 for this cluster. However, we know virtually all clusters in the LMC have line of sight extinction values well below this. The reddening and age are seen as highly correlated using the K-S matching algorithm. This does not happen to the same level when using the $\chi^{2}$ minimization method as shown in Figure \ref{surface_NGC2002.dat_smoothed_SSPPadova.z008.eps}.
Figure \ref{KS_ages} shows the correlation between the ages obtained using the K-S method and the CMD ages. The correlation coefficient is 0.78. The red dashed line is the fit line. For NGC2213 although the CMD log (age/year) is 8.95, the K-S method gives a prediction of 6.8. A closer look at the reddening predicted for this cluster shows a high value of 0.48. For this cluster the age/reddening degeneracy was not resolved properly with the K-S method, this might be due to the bad S/N. We show in Section 4.3 that the difference in age predictions by the K-S test versus the $\chi^{2}$ minimization method vary for S/N $<$ 60 and it is minimum for S/N $>$ 60.

\begin{deluxetable}{|c|c|c|c|c|c|c|c|c|c|c|c|c|c|c|c|}
\tabletypesize{\scriptsize}
\tablecaption{Age predicted by different model libraries using different statistical methods \label{T2}}
\tablewidth{0pt}

\tablehead{

\colhead{Name} & \colhead{Age$^1$} & \colhead{Error} & \colhead{Age$^2$} & \colhead{Error} &  \colhead{Age$^3$} & \colhead{Error} &  \colhead{Age$^4$} & \colhead{Error} & \colhead{Age$^5$} & \colhead{Error} 
}
\startdata

NGC1651  &  8.90  & 60\%  &  8.96 & 54\%  &  8.81 & 68\%  &  9.05 & 44\%  &  8.75 & 72\%  \\
NGC1711  &  7.55  & 41\%  &  7.63 & 70\%  &  7.58 & 51\%  &  -    & -     &  -    & -     \\
NGC1839  &  8.05  & 239\% &  8.11 & 289\% &  8.11 & 289\% &  -    & -     &  -    & -     \\
NGC1850  &  7.75  & 41\%  &  7.70 & 26\%  &  7.76 & 45\%  &  -    & -     &  -    & -     \\
NGC1856  &  8.45  & 255\% &  8.41 & 224\% &  8.36 & 188\% &  8.54 & 337\% &  8.45 & 255\% \\ 
NGC1863  &  7.45  & 51\%  &  7.51 & 44\%  &  7.46 & 50\%  &  -    & -     &  -    & -     \\
NGC1870  &  7.80  & 13\%  &  7.86 & 0\%   &  7.81 & 11\%  &  8.00 & 38\%  &  7.85 & 2\%   \\
NGC1894  &  7.85  & 29\%  &  7.81 & 17\%  &  7.81 & 17\%  &  -    & -     &  -    & -     \\
NGC1903  &  7.85  & 0\%   &  8.16 & 104\% &  7.86 & 2\%   &  8.11 & 82\%  &  8.15 & 100\% \\ 
NGC1983  &  6.65  & 44\%  &  6.74 & 31\%  &  6.46 & 64\%  &  -    & -     &  -    & -     \\
NGC1984  &  6.65  & 37\%  &  6.90 & 12\%  &  6.40 & 65\%  &  -    & -     &  -    & -     \\
NGC1994  &  6.65  & 38\%  &  7.00 & 38\%  &  6.64 & 40\%  &  -    & -     &  -    & -     \\
NGC2002  &  7.00  & 37\%  &  6.82 & 58\%  &  7.00 & 37\%  &  -    & -     &  -    & -     \\
NGC2011  &  6.65  & 26\%  &  6.94 & 45\%  &  6.46 & 52\%  &  -    & -     &  -    & -     \\
NGC2031  &  8.25  & 12\%  &  8.36 & 45\%  &  8.16 & 9\%   &  8.34 & 38\%  &  8.26 & 15\%  \\
NGC2065  &  7.95  & 26\%  &  8.21 & 129\% &  7.96 & 29\%  &  8.20 & 124\% &  8.15 & 100\% \\  
NGC2136  &  7.90  & 100\% &  8.16 & 263\% &  7.91 & 104\% &  -    & -     &  -    & -     \\
NGC2155  &  9.20  & 37\%  &  9.99 & 289\% &  9.16 & 42\%  &  10.1 & 401\% &  9.20 & 37\%  \\ 
NGC2156  &  8.00  & 66\%  &  7.96 & 51\%  &  8.01 & 70\%  &  8.04 & 82\%  &  8.04 & 82\%  \\
NGC2157  &  7.85  & 78\%  &  7.86 & 82\%  &  7.81 & 62\%  &  -    & -     &  -    & -     \\
NGC2164  &  7.90  & 58\%  &  7.86 & 45\%  &  7.91 & 62\%  &  -    & -     &  -    & -     \\
NGC2172  &  7.55  & 41\%  &  7.65 & 26\%  &  7.60 & 34\%  &  7.85 & 17\%  &  7.78 & 0\%   \\
NGC2173  &  9.35  & 7\%   &  9.41 & 23\%  &  9.23 & 19\%  &  9.55 & 70\%  &  9.35 & 7\%   \\
NGC2213  &  6.80  & 99\%  &  9.16 & 62\%  &  6.80 & 99\%  &  9.15 & 58\%  &  7.78 & 93\%  \\
NGC2249  &  8.40  & 62\%  &  8.61 & 38\%  &  8.31 & 69\%  &  8.70 & 24\%  &  8.34 & 67\%  \\
SL234    &  7.80  & 32\%  &  7.81 & 35\%  &  7.81 & 35\%  &  -    & -     &  -    & -     \\
SL237    &  6.90  & 70\%  &  6.90 & 70\%  &  7.26 & 32\%  &  -    & -     &  -    & -     \\

\enddata
\tablecomments{
$^1$ Predicted by  \citet {Delgado05} using K-S test \\
$^2$ Predicted by GALAXEV using the $\chi^{2}$ minimization method \\
$^3$ Predicted by GALAXEV using the K-S test\\
$^4$ Predicted by MILES using the $\chi^{2}$ minimization method for clusters with CMD age equal to or greater than log (Age/year) 7.78 \\ 
$^5$ Predicted by MILES using the K-S test for clusters with CMD age equal to or greater than log (Age/year) 7.78\\
}
\end{deluxetable}

\begin{deluxetable}{|c|c|c|c|c|c|c|c|c|c|c|c|c|c|c|c|}
\tabletypesize{\scriptsize}
\tablecaption{Reddening predicted by different model libraries using different statistical methods \label{T3}}
\tablewidth{0pt}

\tablehead{

\colhead{Name} & \colhead{E(B$-$V)$^1$} & \colhead{Error} & \colhead{E(B$-$V)$^2$} & \colhead{Error} & \colhead{E(B$-$V)$^3$} & \colhead{Error} & \colhead{E(B$-$V)$^4$} & \colhead{Error} & \colhead{E(B$-$V)$^5$} & \colhead{Error}
}
\startdata

NGC1651  &  0.26 &   189\%	&  0.00 &  -100\%   &  0.29 &  222\%   &  0.00 &  -100\% &  0.37 &  311\% \\
NGC1711  &  0.09 &   -44\%	&  0.02 &  -88\%    &  0.05 &  -69\%   &  -    &  -         &  -    &  -   \\
NGC1839  &  0.04 &   -85\%	&  0.00 &  -100\%   &  0.01 &  -96\%   &  -    &  -         &  -    &  -  \\
NGC1850  &  0.11 &   -39\%	&  0.06 &  -67\%    &  0.08 &  -56\%   &  -    &  -         &  -    &  -  \\
NGC1856  &  0.12 &   -43\%	&  0.12 &  -43\%    &  0.15 &  -29\%   &  0.10 &  -52\%  &  0.13 &  -38\%  \\ 
NGC1863  &  0.10 &   -50\%	&  0.05 &  -75\%    &  0.06 &  -70\%   &  -    &  -         &  -    &  -  \\
NGC1870  &  0.06 &   -57\%	&  0.02 &  -86\%    &  0.03 &  -79\%   &  0.01 &  -93\%  &  0.04 &  -71\%  \\
NGC1894  &  0.26 &   160\%	&  0.25 &  150\%    &  0.24 &  140\%   &  -    &  -         &  -    &  -   \\
NGC1903  &  0.17 &   6\%	    &  0.06 &  -63\%    &  0.15 &  -6\%    &  0.06 &  -63\%  &  0.06 &  -63\%  \\ 
NGC1983  &  0.08 &   -11\%	&  0.00 &  -100\%   &  0.22 &  144\%   &  -    &  -         &  -    &  -    \\
NGC1984  &  0.28 &   100\%	&  0.00 &  -100\%   &  0.42 &  200\%   &  -    &  -         &  -    &  -   \\
NGC1994  &  0.20 &   43\%	&  0.05 &  -64\%    &  0.25 &  79\%    &  -    &  -         &  -    &  -   \\
NGC2002  &  0.49 &   308\%	&  0.26 &  117\%    &  0.47 &  292\%   &  -    &  -         &  -    &  -  \\
NGC2011  &  0.26 &   225\%	&  0.00 &  -100\%   &  0.40 &  400\%   &  -    &  -         &  -    &  -   \\
NGC2031  &  0.00 &   -100\%  &  0.00 &  -100\%   &  0.04 &  -56\%   &  0.00 &  -100\% &  0.01 &  -89\%  \\
NGC2065  &  0.15 &   -17\%	&  0.04 &  -78\%    &  0.13 &  -28\%   &  0.04 &  -78\%  &  0.06 &  -67\%  \\  
NGC2136  &  0.13 &   30\%	&  0.05 &  -50\%    &  0.11 &  10\%    &  -    &  -         &  -    &  -   \\
NGC2155  &  0.00 &   -100\%  &  0.00 &  -100\%   &  0.00 &  -100\%   &  0.00 &  -100\% &  0.01 &  -50\% \\ 
NGC2156  &  0.03 &   -70\%	&  0.00 &  -100\%   &  0.02 &  -80\%   &  0.00 &  -100   &  0.03 &  -70\% \\
NGC2157  &  0.16 &   60\%	&  0.14 &  40\%	   &  0.15 &  50\%    &  -    &  -         &  -    &  -   \\
NGC2164  &  0.01 &   -90\%	&  0.00 &  -100\%   &  0.00 &  -100\%   &  -    &  -         &  -    &  -   \\
NGC2172  &  0.12 &   20\%	&  0.04 &  -60\%    &  0.07 &  -30\%   &  0.05 &  -50\%  &  0.05 &  -50\%  \\
NGC2173  &  0.00 &   -100\%  &  0.00 &  -100\%   &  0.00 &  -100\%   &  0.00 &  -100\% &  0.00 &  -100\% \\
NGC2213  &  0.48 &   433\%	&  0.00 &  -100\%   &  0.49 &  444\%   &  0.00 &  -100\% &  0.47 &  422\%   \\
NGC2249  &  0.12 &   1100\%  &  0.00 &  -100\%   &  0.16 &  1500\%   &  0.00 &  -100\% &  0.14 &  1300\%  \\
SL234    &  0.04 &   -73\%	&  0.01 &  -93\%    &  0.01 &  -93\%   &  -    &  -         &  -    &  -   \\
SL237    &  0.26 &   53\%	&  0.23 &  35\%	   &  0.34 &  100\%   &  -    &  -         &  -    &  -  \\
\enddata
\tablecomments{
$^1$ Predicted by  \citet {Delgado05} using K-S test \\
$^2$ Predicted by GALAXEV using the $\chi^{2}$ minimization method \\
$^3$ Predicted by GALAXEV using the K-S test\\
$^4$ Predicted by MILES using the $\chi^{2}$ minimization method for clusters with CMD age equal to or greater than log (Age/year) 7.78 \\ 
$^5$ Predicted by MILES using the K-S test for clusters with CMD age equal to or greater than log (Age/year) 7.78\\
}
\end{deluxetable}

\section{Stellar Populations Model Libraries}

The two new models added to ASAD$_2$ are GALAXEV \citep{Bruzual03} and MILES \citep{Miles2010} as recently updated in \citet{Miles2015}

\subsection{GALAXEV}   

We use the optical range of the GALAXEV \citep{Bruzual03} models which contain the spectral evolution of stellar populations at a resolution of 3$\rm\AA$ (FWHM). We chose the spectral models derived using the Padova 1994 \citep{Bertelli94} evolutionary tracks and the Salpeter (1955) IMF with lower mass cutoff 0.1 solar mass and upper mass cutoff of 100 solar mass.

The ages are converted into log (age/year) and rounded to two decimal points\footnote {The ages provided by the model are not perfectly uniform in the step size. They start at log (age/year) 5.10, and increase in step of 0.05 up to 6.00 then increase in steps of 0.02 up to 7.48 then vary slightly in the step size up to 10.10. The spectral fluxes between log (age/year) 5.1 and 6.2 are identical so ASAD$_2$ skips the ages less than log (age/year) 6.2.}.

\begin{figure}
\includegraphics[angle=270,scale=0.3]{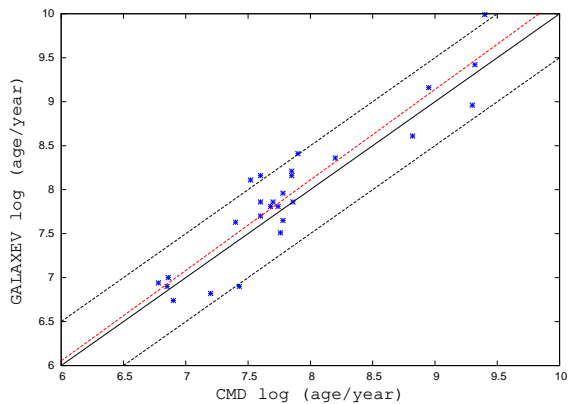}
\caption{The correlation between the ages obtained using GALAXEV with the $\chi^{2}$ minimization method versus the CMD ages. The correlation coefficient is 0.93. The red dashed line is the fit line. The dashed lines represent the upper and lower limit of the range of ages within log (age/year) 0.5.}
\label{GALAXEVvsCMD}
\end{figure}

\begin{figure}
\includegraphics[angle=270,scale=0.3]{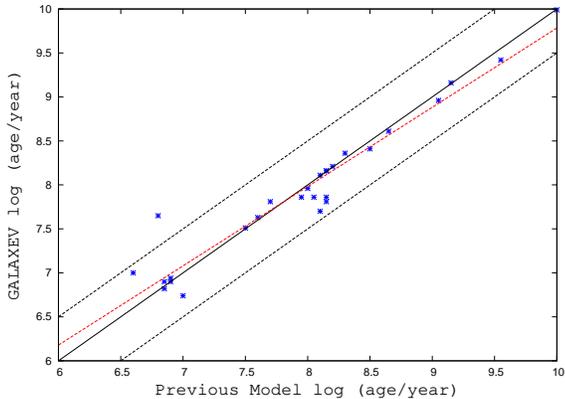}
\caption{The correlation between the ages obtained using GALAXEV with the $\chi^{2}$ minimization method versus the ages obtained using the model of \citet {Delgado05}. The correlation coefficient is 0.96. The red dashed line is the fit line. The dashed lines represent the upper and lower limit of the range of ages within log (age/year) 0.5.}
\label{GALAXEVvsOLD}
\end{figure}

We used fixed metallicity Z = 0.008\footnote {represented by m52 in the model library}. The results obtained using the $\chi^{2}$ minimization method and the percentage errors are listed in columns 4 and 5 of Tables \ref{T2} and \ref{T3}. Figure \ref{GALAXEVvsCMD} shows the correlation between the ages obtained using GALAXEV using the $\chi^{2}$ minimization method versus the CMD ages. The correlation coefficient is 0.93. Figure \ref{GALAXEVvsOLD} shows the correlation between the ages obtained using GALAXEV with the $\chi^{2}$ minimization method versus the ages obtained using the model of \citet {Delgado05}. The correlation coefficient is 0.96. The difference in the predicted log (age/year) by the two models for 50\% of the clusters is less than 0.05. We expect the deviating clusters at around log (age/year) 6.7 and 8.1 to correspond to differences in the treatment of these models of the Red Supergiants phase, and the onset of the AGB, respectively.

\begin{figure}
\includegraphics[angle=270,scale=0.3]{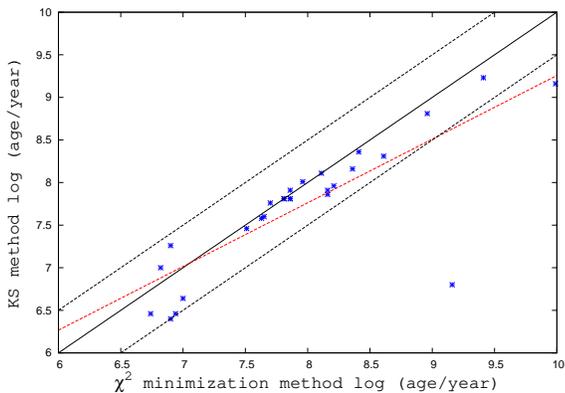}
\caption{The correlation between the ages obtained using the K-S test versus the ages obtained using the $\chi^{2}$ minimization method with GALAXEV model. The correlation coefficient is 0.81. The red dashed line is the fit line. The dashed lines represent the upper and lower limit of the range of ages within log (age/year) 0.5.} 
\label{chiVSks}
\end{figure}

The results obtained using the K-S method with the GALAXEV model and the percentage errors are listed in columns 6 and 7 of Tables \ref{T2} and \ref{T3}. Figure \ref{chiVSks} shows the correlation between the ages obtained using the K-S test versus the ages obtained using the $\chi^{2}$ minimization method. The correlation coefficient is 0.81.

The outlier is NGC2213, when removed from the calculations the correlation coefficient is 0.94. For this cluster, the $\chi^{2}$ minimization method predicts an old age with zero reddening, while the K-S test predicts a young age with a high reddening. Comparing the predicted ages with the CMD age, we find that the $\chi^{2}$ minimization method predicts a more accurate result for this cluster.

\begin{figure}
\includegraphics[angle=0,scale=0.7]{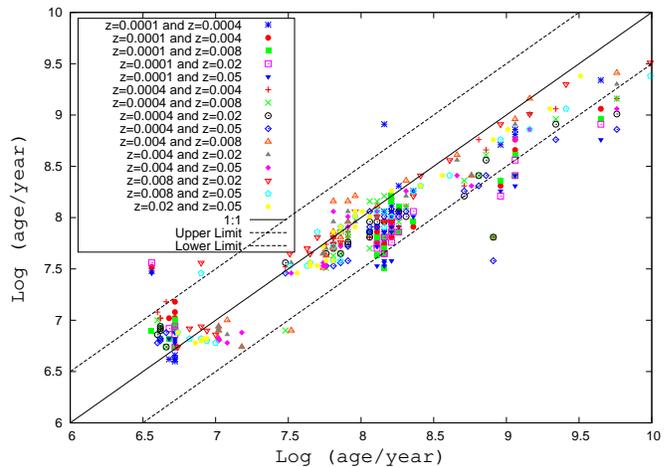}
\caption{Age prediction using different combinations of metallicity as indicated in the key. See the text for more details.}
\label{GALAXEVMetal}
\end{figure}

To test the effect of metallicity on our method, we used the different metallicities provided by this model, to compare the ages obtained by each metallicity. Figure \ref{GALAXEVMetal} shows the age prediction using different combinations of metallicity as indicated in the key. The blue stars show log (age/year) obtained using metallicity Z= 0.0001 versus log (age/year) obtained using metallicity Z= 0.0004. The red circles show log (age/year) obtained using metallicity Z= 0.0001 versus log (age/year) obtained using metallicity Z= 0.004. The green squares show log (age/year) obtained using metallicity Z= 0.0001 versus log (age/year) obtained using metallicity Z= 0.008 and so on. The dashed lines represent the upper and lower limit of the range of ages within log (age/year) 0.5. 369 values out of 405 lie within that range, that is 91\%. Few outliers are noted for the young clusters. Most outliers are for the older clusters (log (age/year) $>$ 9) where the age/metallicity degeneracy is noticeable. For the oldest ages (log (age/year) $>$ 9.5) most points are outliers which mean that our technique is not very suitable for these old clusters.
Our method is applicable to the young (age/year) $<$ 9) clusters but not appropriate as the age/metallicity degeneracy becomes too relevant, preventing us to assume a metallicity.

\subsection{MILES}  

MILES website allows choosing the preferred model configuration. Any configuration desired can be easily imported into ASAD$_2$. We chose the models that employ the \citet {Girardi2000} theoretical isochrones (Padova00) and Salpeter (1955) IMF converted to the observational plane on the basis of extensive stellar photometric libraries and the MILES stellar spectral library. A particularly important peculiarity of the MILES spectra for this work is its excellent flux-calibration quality and good parameters coverage. ASAD$_2$ first groups models with the same metallicity together, then extracts the flux and stores it for the corresponding wavelength, one flux column for each age\footnote {In the MILES model, the flux values are divided into separate files based on the model's metallicity, age, and IMF slope. The values of the metallicity, age, and IMF slope are encoded in the name of each file}.

The ages are converted into log (age/year) and rounded to two decimal points. Note that the ages provided by the model start at log (age/year) 7.78, which is relatively large compared to the other models. Another option is to start with 7.4 when using the model version based on BaSTI isochrones. We chose the Padova library for uniformity (with the other models used in ASAD$_2$). The ages increase in step size of roughly 0.05\footnote {For the LMC clusters we used the fixed metallicity Z = 0.008 (represented by [M/H] = - 0.4 in the model library)}. The results and the percentage errors are listed in columns 8 and 9 of Tables \ref{T2} and \ref{T3} for the $\chi^{2}$ minimization method.

\begin{figure}
\includegraphics[angle=270,scale=0.3]{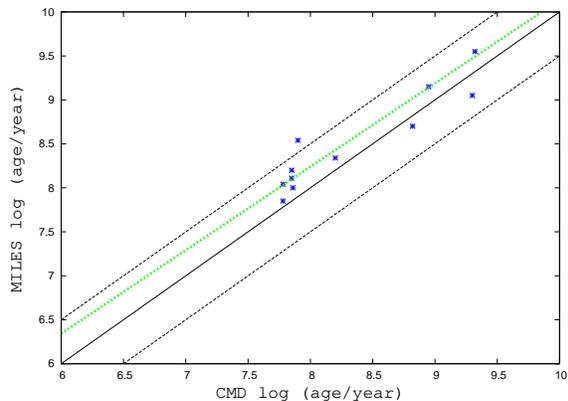}
\caption{The correlation between the ages obtained using MILES with the $\chi^{2}$ minimization method versus the CMD ages when excluding the clusters younger than log (age/year) of 7.78. The green dotted line is the fit line.}
\label{MILESvsCMDNew}
\end{figure}

\begin{figure}
\includegraphics[angle=270,scale=0.3]{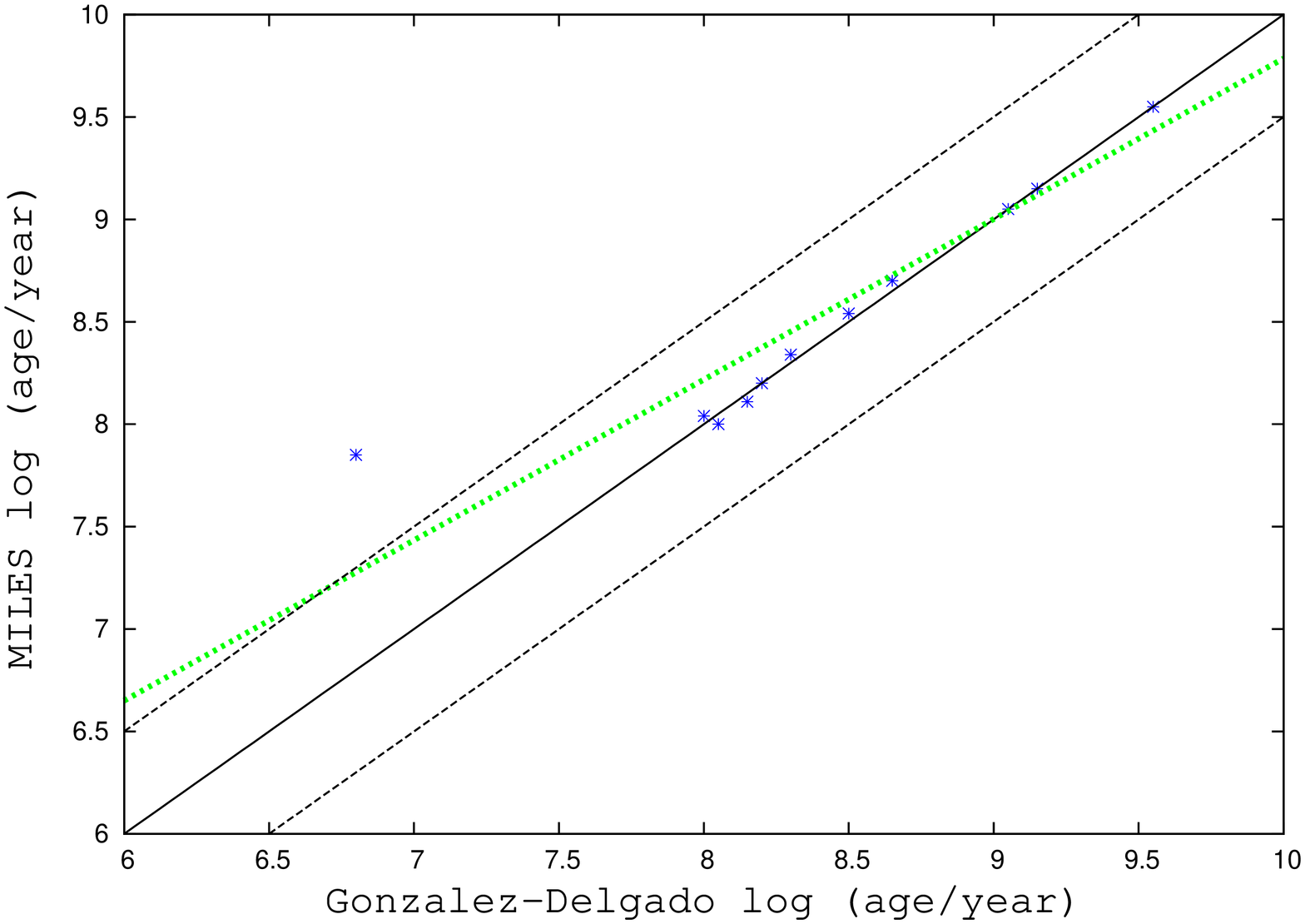}
\caption{The correlation between the ages obtained using MILES with the the $\chi^{2}$ minimization method versus the ages obtained using the model of \citet {Delgado05} when excluding the clusters younger than log (age/year) of 7.78. The green dotted line is the fit line.}
\label{MILESvsOLDNew}
\end{figure}

\begin{figure}
\includegraphics[angle=0,scale=0.2]{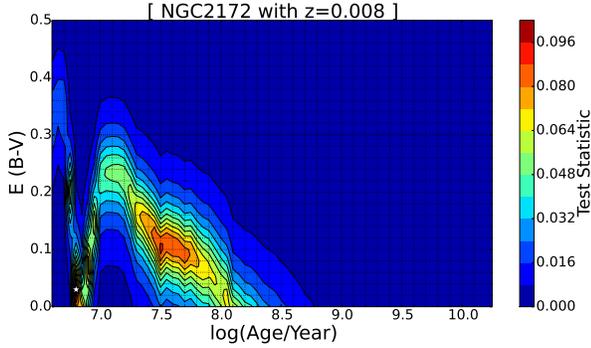}
\caption{The surface plot of NGC2172 predicted by \citet {Delgado05} model with the $\chi^{2}$ minimization method. A second possible solution is noticed around log (age/year) 7.5}
\label{NGC2172}
\end{figure}

\begin{figure}
\includegraphics[angle=270,scale=0.3]{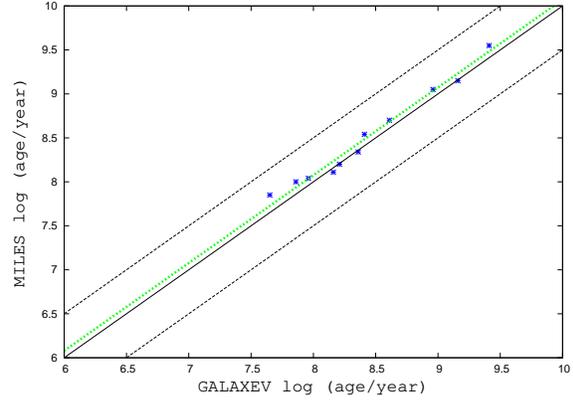}
\caption{The correlation between the ages obtained using MILES versus the ages obtained using GALAXEV when using the $\chi^{2}$ minimization method when excluding the clusters younger than log (age/year) of 7.78. The green dotted line is the fit line.}
\label{MILESvsGALAXEVNew}
\end{figure}

Figure \ref{MILESvsCMDNew} shows the correlation between the ages obtained using MILES with the the $\chi^{2}$ minimization method versus the CMD ages. We excluded the clusters with a CMD age younger than log (age/year) 7.78 (12 clusters of our sample). The correlation coefficient is 0.92. Figure \ref{MILESvsOLDNew} shows the correlation between the ages obtained using MILES versus the ages obtained using the model of \citet {Delgado05}. The outlier is NGC2172. \citet {Delgado05} predicts a log (age/year) 6.8 while MILES predict a log (age/year) 7.85. MILES prediction is closer to the CMD age of this cluster (log (age/year) = 7.78). Figure \ref{NGC2172} shows that there is another close possible solution around log (age/year) 7.5. The outlier in Figures \ref{MILESvsOLDNew} shows that MILES chooses the older option among the two possible ones, which cause the two models to disagree for this particular cluster.
Figure \ref{MILESvsGALAXEVNew} shows the correlation between the ages obtained using MILES versus the ages obtained using GALAXEV. The correlation coefficient is 0.99 when excluding the young clusters.

\begin{figure}
\includegraphics[angle=270,scale=0.3]{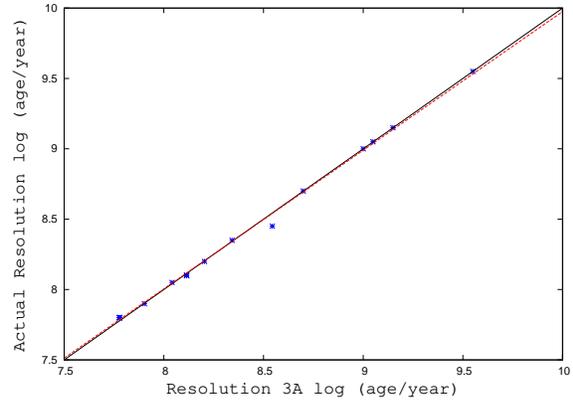}
\caption{The results obtained using a fixed resolution for the model higher than that of the data, compared to the results obtained when matching the resolution of the model to that of the data.}
\label{CompareResolution}
\end{figure}

It is worth mentioning here that the spectral resolution of the model is greater than that of the clusters. We used a resolution of 3$\rm\AA$ for the model to make it similar to the resolution used with the previous models (to make the comparison consistent). To test our results, we did the fits again using the resolution of the model that matches the data (3.6$\rm\AA$ for SOAR data and 14$\rm\AA$ for Blanco data as described in \citet{Asad14}), the results are almost identical as shown in Figure \ref{CompareResolution}. The outlier is NGC1856 observed with Blanco. 

\begin{figure}
\includegraphics[angle=270,scale=0.3]{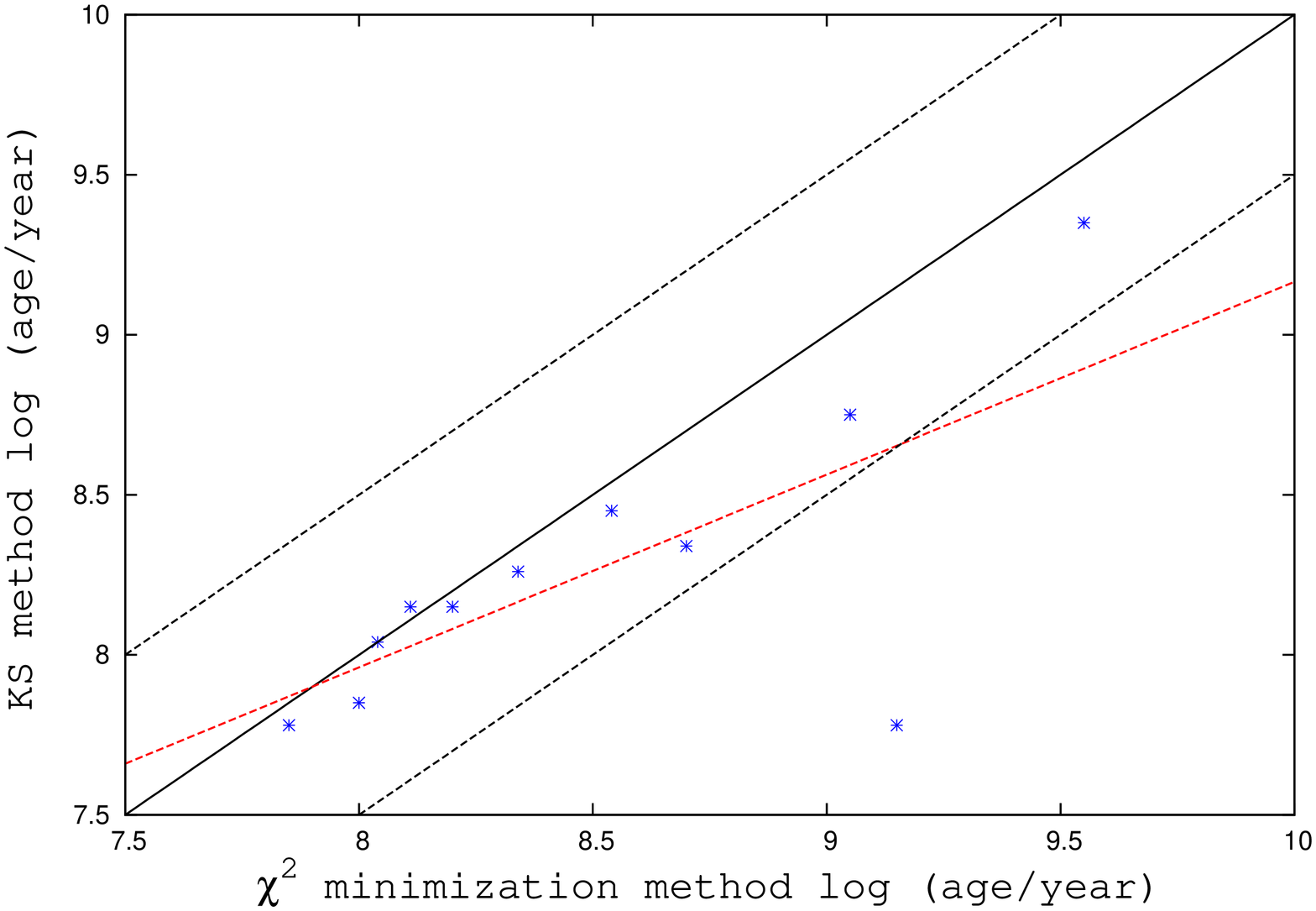}
\caption{The correlation between the ages obtained using the K-S test versus the ages obtained using the $\chi^{2}$ minimization method for MILES model. The correlation coefficient is 0.80. The red dashed line is the fit line.}
\label{chiVSksMILES}
\end{figure}

The results obtained with MILES using the K-S test and the percentage errors are listed in columns 10 and 11 of Tables \ref{T2} and \ref{T3}.
 Figure \ref{chiVSksMILES} shows the correlation between the ages obtained using the K-S test versus the ages obtained using the $\chi^{2}$ minimization method. The correlation coefficient is 0.80. NGC2213 is an outlier. When compared with the CMD ages the $\chi^{2}$ minimization method gives a better prediction.

\begin{figure}
\includegraphics[angle=0,scale=0.7]{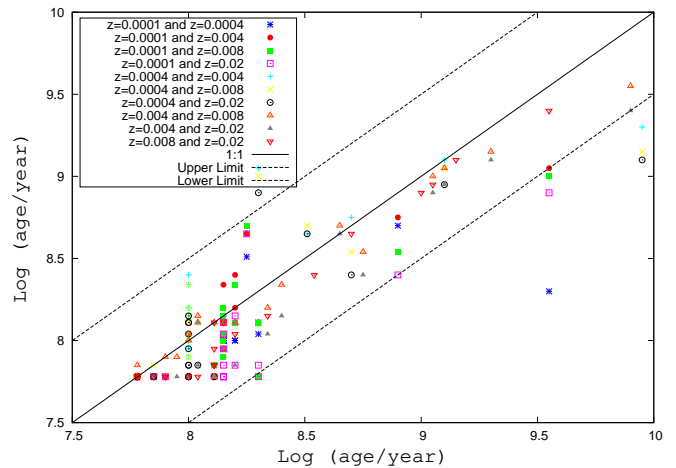}
\caption{Age prediction using different combinations of metallicity for the MILES model as indicated the key. See the text for more details.}
\label{milesMetal}
\end{figure}

As we did with GALAXEV, we use the different metallicities of MILES to compare the ages obtained by each metallicity. Figure \ref{milesMetal} shows the age prediction using different combinations of metallicity as indicated the in key of the figure.  

The blue stars show log (age/year) obtained using metallicity Z= 0.0001 versus log (age/year) obtained using metallicity Z= 0.0004. The red circles show log (age/year) obtained using metallicity Z= 0.0001 versus log (age/year) obtained using metallicity Z= 0.004. The green squares show log (age/year) obtained using metallicity Z= 0.0001 versus log (age/year) obtained using metallicity Z= 0.008 and so on. The dashed lines represent the upper and lower limit of the range of ages within 0.5 log (Age/year). 261 values out of 270 lie within that range, that is 96.7\%. We conclude that metallicity does not strongly affect the age determination for this method for log (age/year) $<$ 9.5. 

\begin{figure}
\includegraphics[angle=270,scale=0.3]{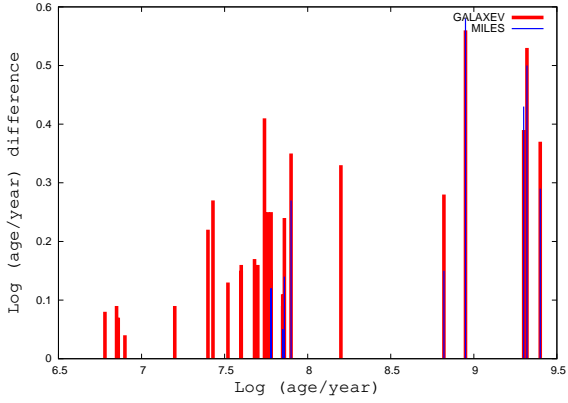}
\caption{The average absolute value of the difference in log (age/year) obtained using the different metallicities of both GALAXEV and MILES.}
\label{Boxes.eps}
\end{figure}

Figure \ref{Boxes.eps} shows the average absolute values of the difference in log (age/year) obtained using the different metallicities of both GALAXEV and MILES. The figure shows three main ranges. For young cluster (log (Age/year) $<$ 7.4) the average age difference is less than 0.1. For intermediate ages (7.4 $<$ log (Age/year) $<$ 8.8) the average difference is around 0.2. Finally for ages $>$ 8.8 the average difference is $>$ 0.35. Except for two outliers, the average difference in log (Age/year) obtained by different metallicities with MILES is less than that seen for GALAXEV. 

\subsection{Dependance on Age, S/N and Resolution}
To better understand the difference in the results obtained by the two statistical results for both GALAXEV and MILES we investigate the dependance of this difference on three factors: age, S/N and resolution. 

\begin{figure}
\includegraphics[angle=270,scale=0.3]{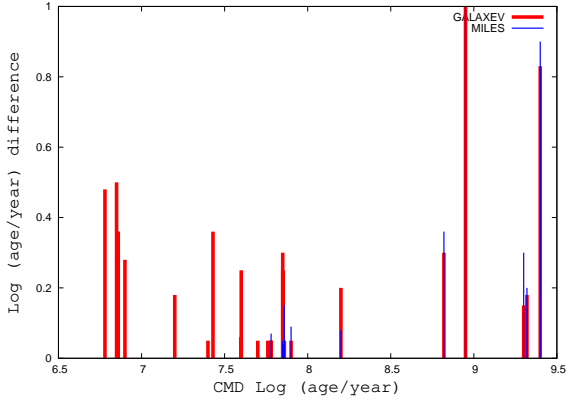}
\caption{The absolute values of the difference in log (age/year) obtained using the two statistical methods (KS method - $\chi^{2}$ minimization method) versus CMD log (age/year).}
\label{BoxesMethods.eps}
\end{figure}

Figure \ref{BoxesMethods.eps} shows the absolute values of the difference in log (age/year) obtained using the two statistical methods (KS method - $\chi^{2}$ minimization method) versus CMD log (age/year). The least difference is seen for intermediate ages 7 $<$ log (age/year) $<$ 8.5. 

\begin{figure}
\includegraphics[angle=270,scale=0.3]{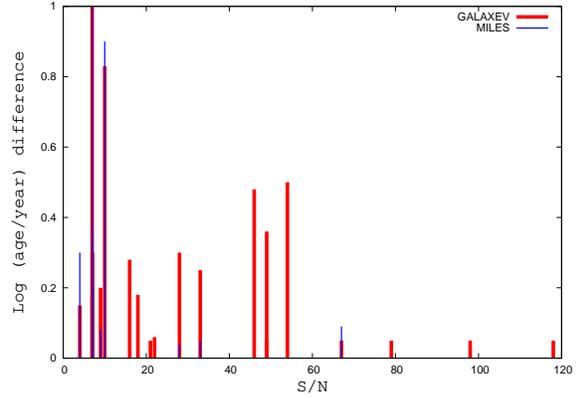}
\caption{The absolute values of the difference in log (age/year) obtained using the two statistical methods (KS method - $\chi^{2}$ minimization method) for versus S/N.}
\label{BoxesSN.eps}
\end{figure}

Figure \ref{BoxesSN.eps} shows the absolute values of the difference in log (age/year) obtained using the two statistical methods (KS method - $\chi^{2}$ minimization method) versus S/N. The difference vary for S/N $<$ 60 and it is minimum for S/N $>$ 60.

\begin{figure}
\includegraphics[angle=270,scale=0.3]{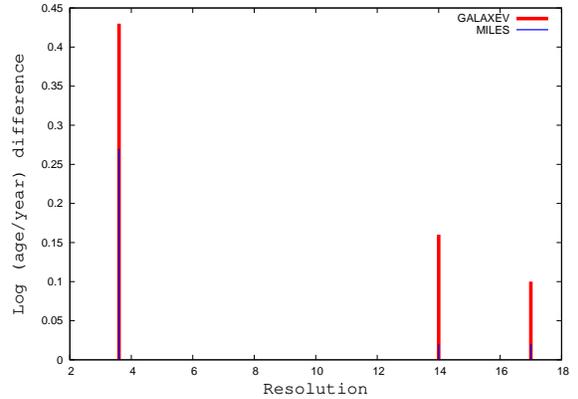}
\caption{The absolute values of the difference in log (age/year) obtained using the two statistical methods (KS method - $\chi^{2}$ minimization method) versus resolution (FWHM) available.}
\label{BoxesRes.eps}
\end{figure}

Figure \ref{BoxesRes.eps} shows the average absolute values of the difference in log (age/year) obtained using the two statistical methods (KS method - $\chi^{2}$ minimization method) versus resolution element in angstroms (FWHM) available. For GALAXEV the difference decreases as the resolution increase. For MILES the difference is the same for the two resolutions 14$\rm\AA$ (FWHM) and 17$\rm\AA$ (FWHM).

\subsection {Error Analysis}

It is noted from the age values of Table 2 that the ages predicted by the two models (GALAXEV and MILES) when using either the $\chi^{2}$ minimization method and the K-S test are mostly within a range of log (age/year) 0.5. As a means to estimate the uncertainty in our aging method, we have derived the relative error for each stellar cluster in our sample using the equation:

\begin{equation}  Error = (\frac{|(Age_{CMD}) - (Age_{predicted})|}{(Age_{CMD})}).
\end{equation}

Here $Age_{ CMD}$ is the CMD age from the literature in years, $Age_{predicted}$ is the age obtained in this work in years. These values are presented for each cluster in Table \ref{T2}. There are few age predictions with percentage error greater than $100\%$.
For GALAXEV age predictions, there are three clusters that have percentage error greater than $100\%$ regardless of the statistical method used, and three clusters for which the percentage error is greater than $100\%$ for the age predicted by the $\chi^{2}$ minimization method, while the age predicted by the K-S test is closer to the correct value. Overall ASAD$_2$ can predict good age estimates for unresolved clusters.

\section {Reddening}

\begin{figure}
\includegraphics[angle=270,scale=0.3]{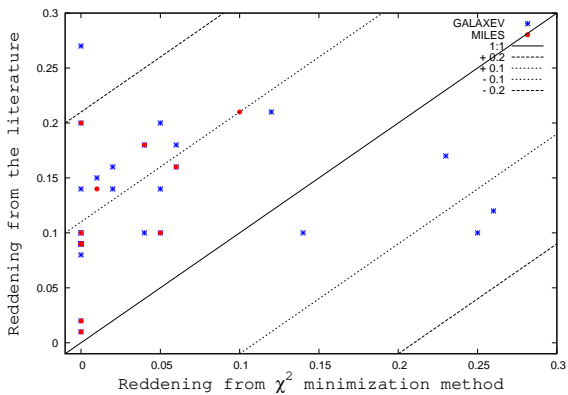}
\caption{The correlation between the reddening obtained using the $\chi^{2}$ minimization method with both GALAXEV and MILES models versus the literature values. Note that only clusters older than log (age/year) 7.78 were included for MILES.}
\label{chiREDnoYOUNG}
\end{figure}

The main goal of the paper is to analyze the age determination for different models using different statistical methods. Although age and reddening are not physically related, the shape of the integrated spectrum for a cluster depends on both age and reddening. It is noted from Table \ref{T3} that the predicted reddening values when using the $\chi^{2}$ minimization method are within the range found in the literature for resolved clusters (i.e: $<$ 0.35), however the K-S test can predict E(B$-$V) values as high as 0.49 when constraining the upper limit to 0.5. Greater values are obtained when increasing the upper limit (see Section 2 for a discussion on NGC2002). Determining the reddening is not easy for unresolved clusters. In this section we analyze the reddening predictions obtained by ASAD$_2$ using the $\chi^{2}$ minimization method as shown in Figure \ref{chiREDnoYOUNG}. 
The percentage error for the reddening determination is listed in Table \ref{T3}. We used the equation:
\begin{equation}  Error = (\frac{|(E(B-V)_{Lit.}) - (E(B-V)_{pred.})|}{(E(B-V)_{Lit.})}).
\end{equation}
Where E(B-V)$_{Lit.}$ refers to the literature value and E(B-V)$_{pred.}$ refers to the predicted value in this work.

\begin{figure}
\includegraphics[angle=270,scale=0.3]{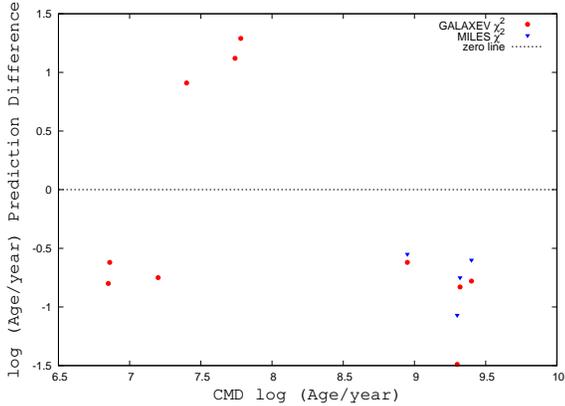}
\caption{The clusters that have a difference in the age prediction between the two spectral ranges larger than 0.5 as a function of CMD age.For MILES we excluded the clusters with CMD age younger than log (age/year) 7.78}
\label{AgeDiffChi}
\end{figure}

\begin{figure}
\includegraphics[angle=270,scale=0.3]{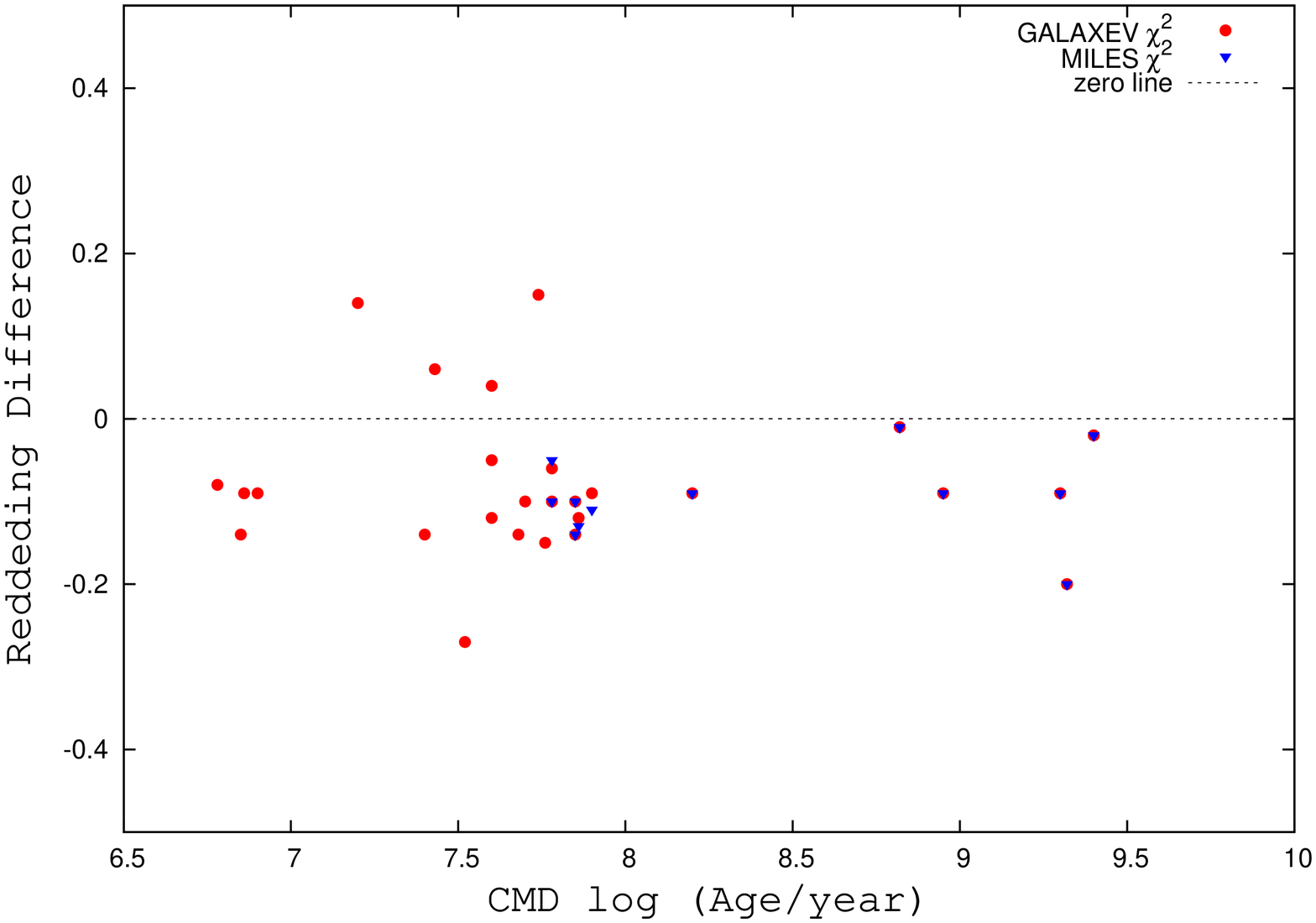}
\caption{The difference in reddening values (our predicted values - literature values) versus the CMD age. For MILES we excluded the clusters with CMD age younger than log (age/year) 7.78}
\label{RedDiffChiNew}
\end{figure}

To understand the correlation between the reddening prediction as a function of age, we plot in Figure \ref{RedDiffChiNew} the difference in reddening values (our predicted values - literature values) versus the CMD age. For MILES we excluded the clusters with CMD age younger than log (age/year) 7.78. Both Figures \ref{chiREDnoYOUNG} and \ref{RedDiffChiNew} show that our method underestimates the values of the reddening except for 4 intermediate-age clusters. 
NGC2002 has a CMD log (age/year) 7.20 and a literature reddening 0.12 while GALAXEV predicts a log (age/year) 6.82 with a reddening of 0.26. SL237 has a CMD log (age/year) 7.43 with a literature reddening 0.17 while GALAXEV predicts a log (age/year) 6.90 with a reddening 0.23. For such young age regime the rapidly varying spectrum shape as a result of the supergiants contributions around log (age/year) 7.0 is scattering both the age and reddening estimates. This effect should be better seen in the red part of the spectrum. We examined this by obtaining the age estimates when using the blue part of our spectrum (3626 $-$ 4700 $\rm\AA$) and the red part of the spectrum (4700 $-$ 6230 $\rm\AA$)) separately. Figure \ref{AgeDiffChi} shows the clusters that have a difference in the age prediction between the two spectral ranges larger than 0.5 as a function of CMD age. The clusters showing the largest age difference are located around the supergiants contribution, at either side of the AGB peak around log (age/year) 9.0 and around log (age/year) 7.8. We can conclude that the sharp spectrum transition originated at these ages are limiting our ability to provide values in agreement with the CMD estimates and as a result the reddening determination is not accurate.

\section {Discussion on Specific Clusters}

In this section we discuss the details of four specific clusters of different ages, ranging from CMD log (age/year) 6.86 to 9.32. 
The full images are available as supplementary online material with this paper.

\subsection {NGC1994}

\begin{figure}
\begin{tabular}{cc}
\resizebox{75mm}{!}{\includegraphics{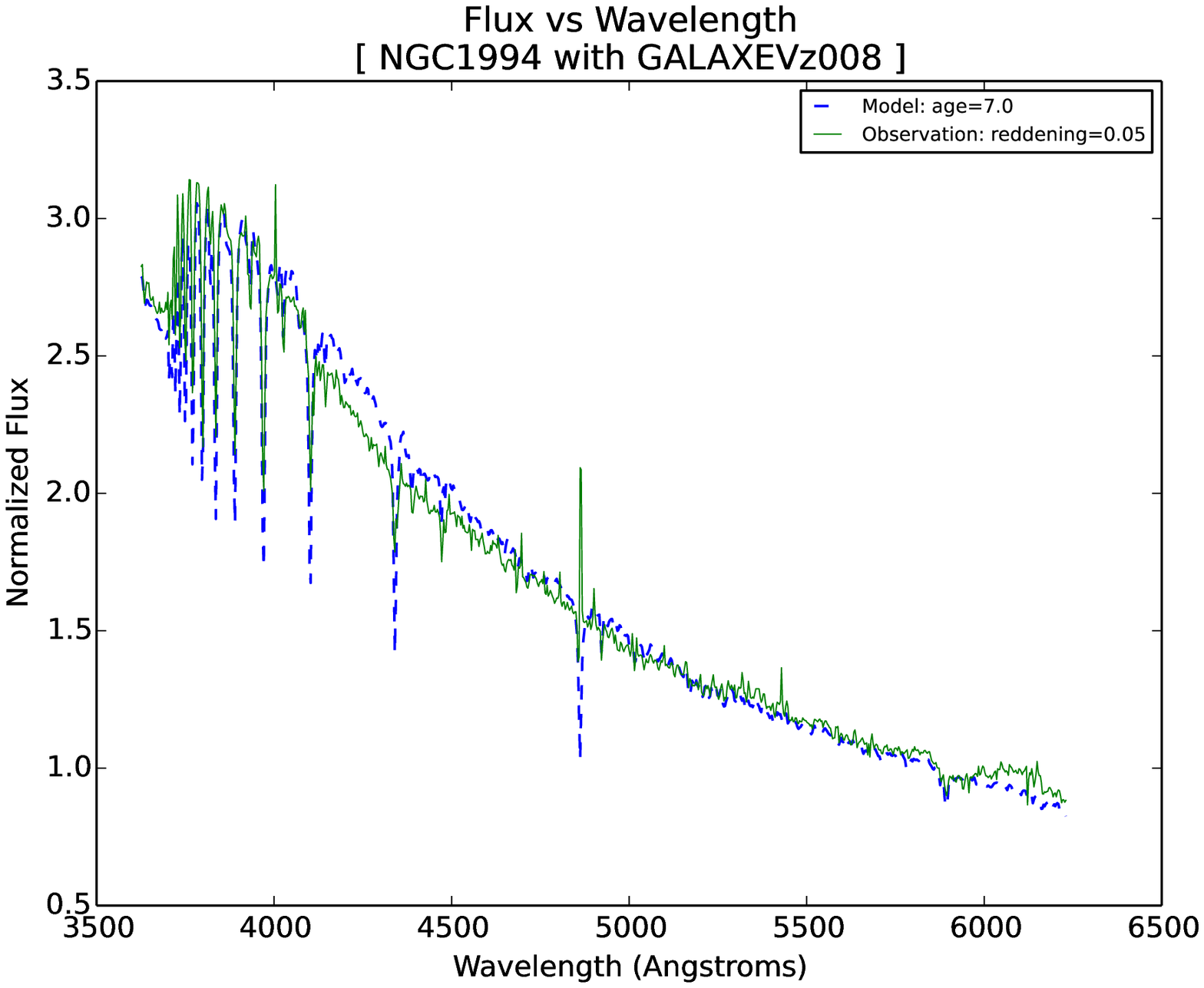}} &
\resizebox{75mm}{!}{\includegraphics{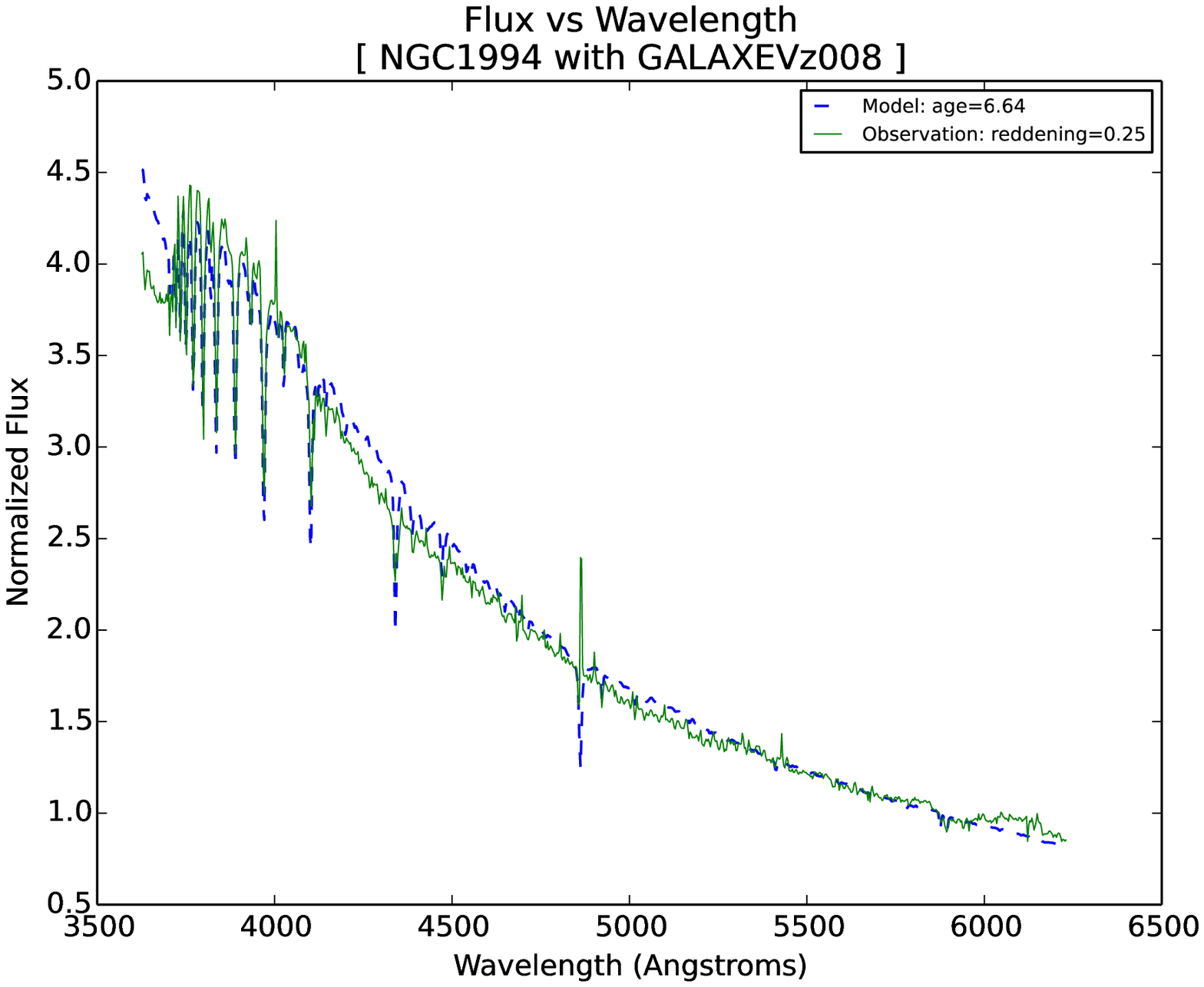}} \\
\resizebox{75mm}{!}{\includegraphics{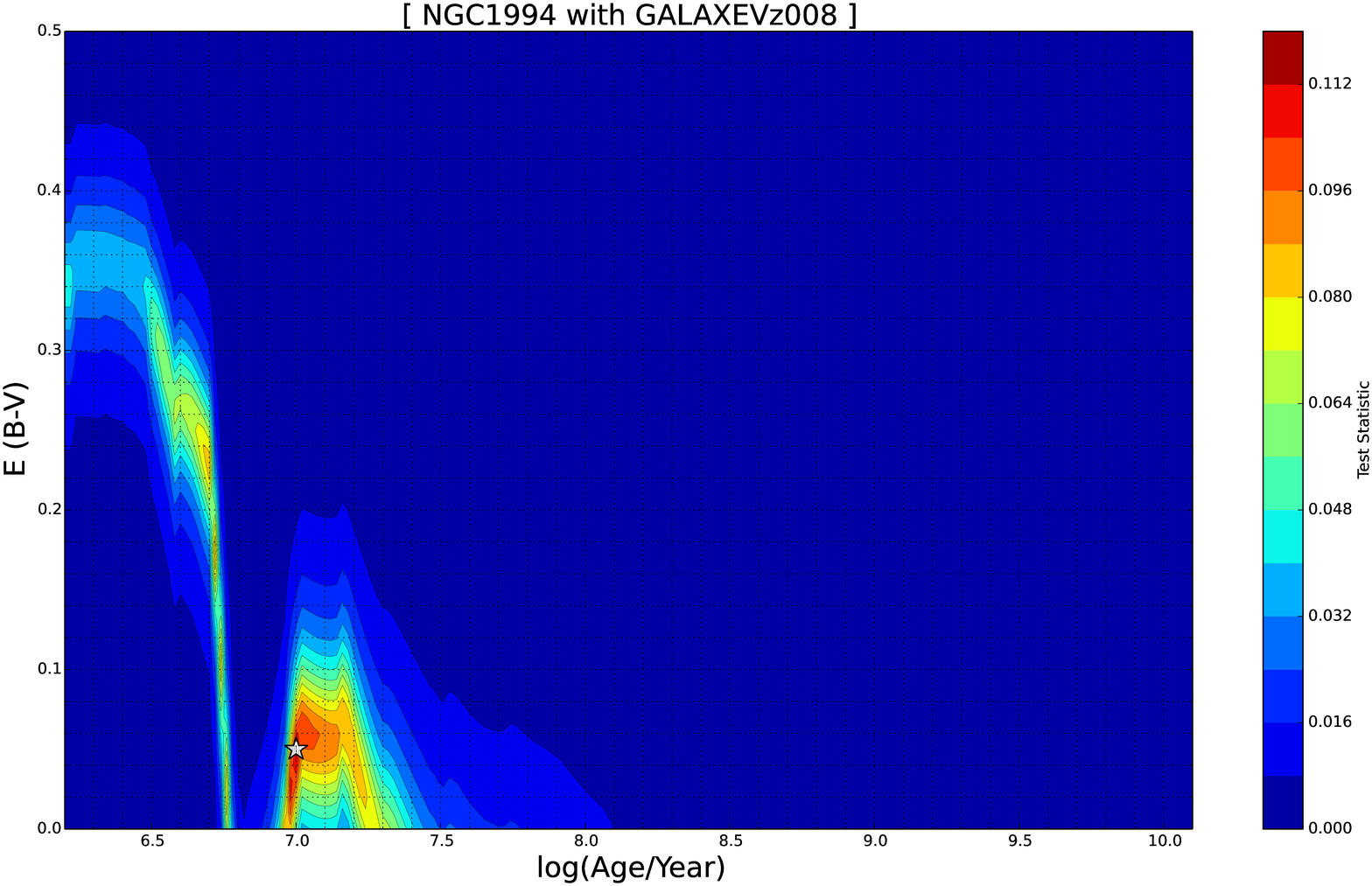}} &
\resizebox{75mm}{!}{\includegraphics{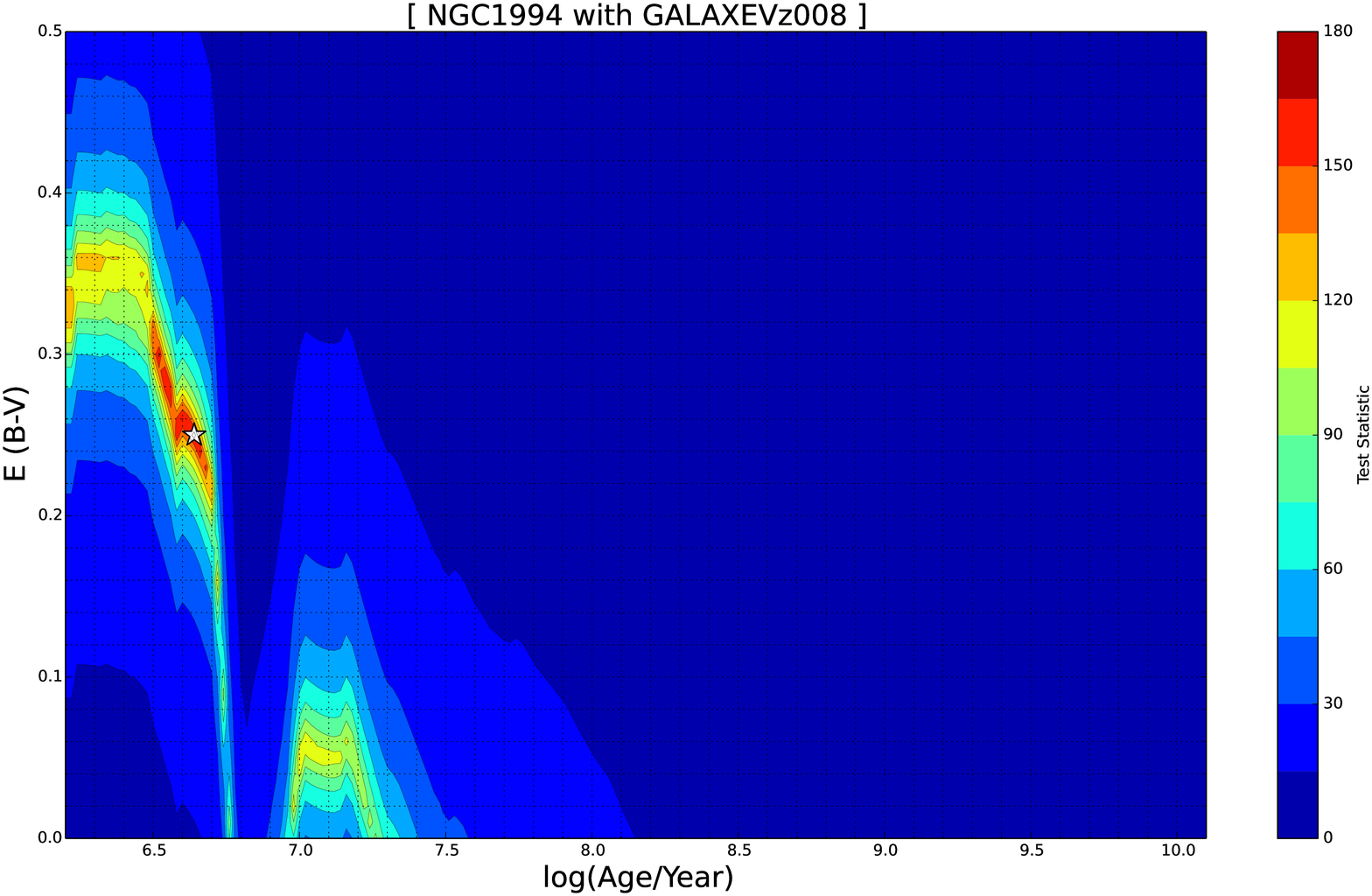}} \\

\end{tabular}
\caption{Results for NGC1994. The left column shows the results of the $\chi^{2}$ minimization method and the right column shows the results of the K-S test.}
\label{NGC1994spectra}
\end{figure}

NGC1994 is a young cluster with a CMD log (age/year) 6.86. Figure \ref{NGC1994spectra} shows the plots obtained by ASAD$_2$ for this cluster. The left column shows the results of the $\chi^{2}$ minimization method and the right column shows the results of the K-S test. 
The curves show the match between the dereddened observed integrated spectrum of NGC1994 and the best GALAXEV model. We notice an offset between 4100$\rm\AA$ and 4300$\rm\AA$ for both statistical methods.
The surface plots represent the inverse of all the possible solution for age-reddening combination. We notice that the results of GALAXEV show a range of possible solutions for the reddening (with log (age/year) of 7) when using the $\chi^{2}$ minimization method, and a range of close solutions between reddening of 0.23 and 0.30 and log (age/year) 6.5 and 6.7.
As discussed in the previous section, we do not show the results from the MILES models as they cannot predict such young ages. 

\subsection {NGC2002}

\begin{figure}
\begin{tabular}{cc}
\resizebox{75mm}{!}{\includegraphics{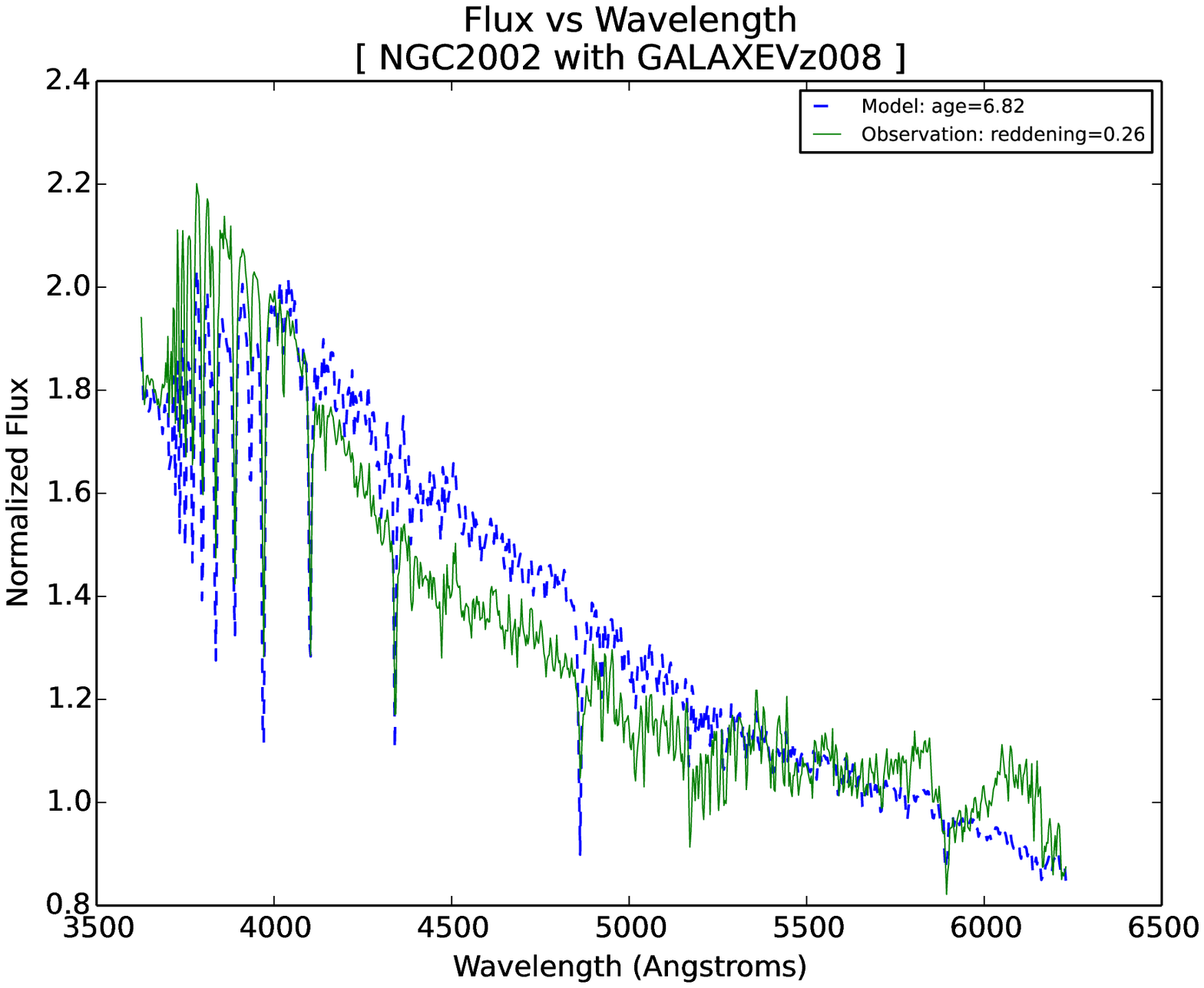}} &
\resizebox{75mm}{!}{\includegraphics{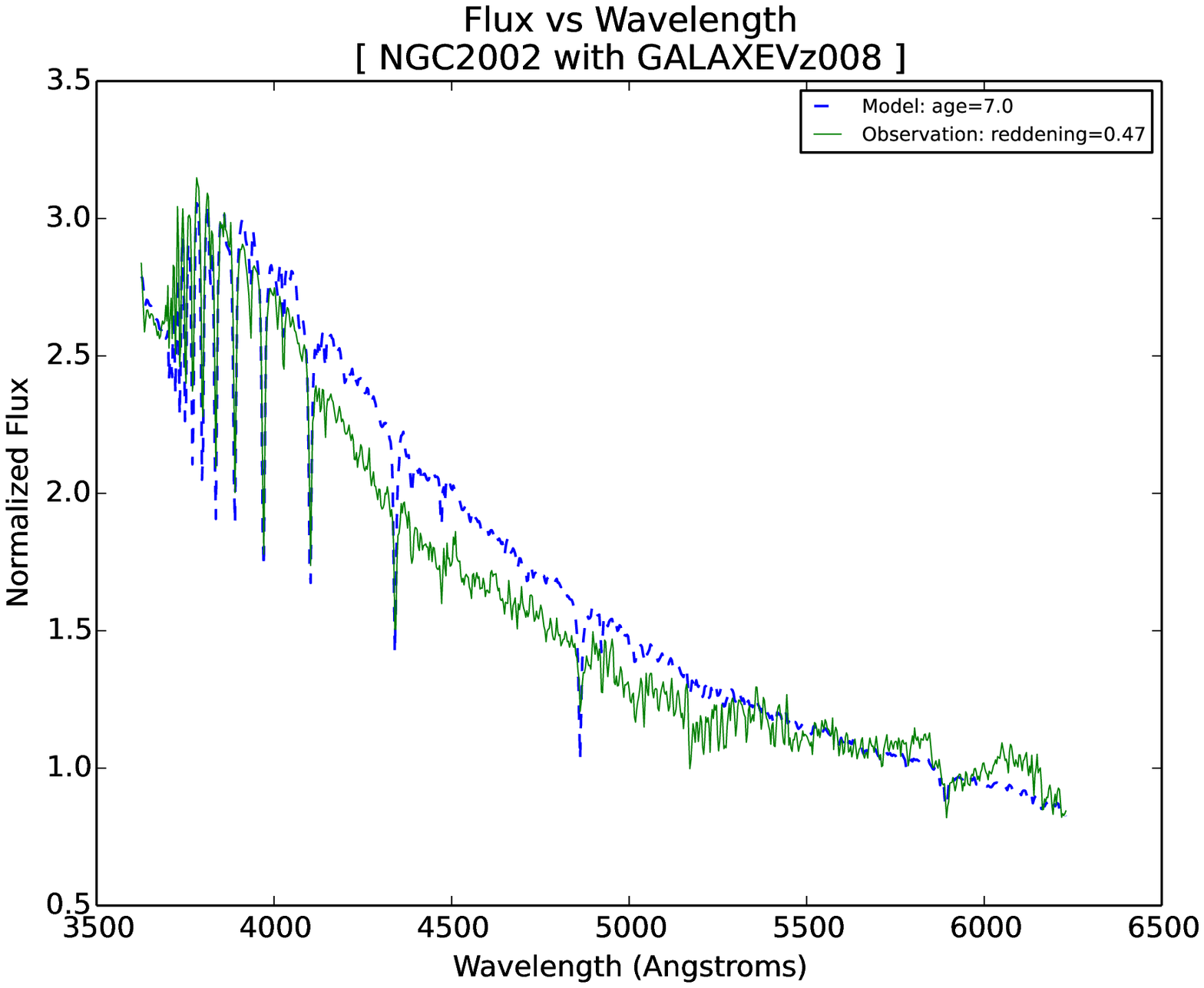}} \\
\resizebox{75mm}{!}{\includegraphics{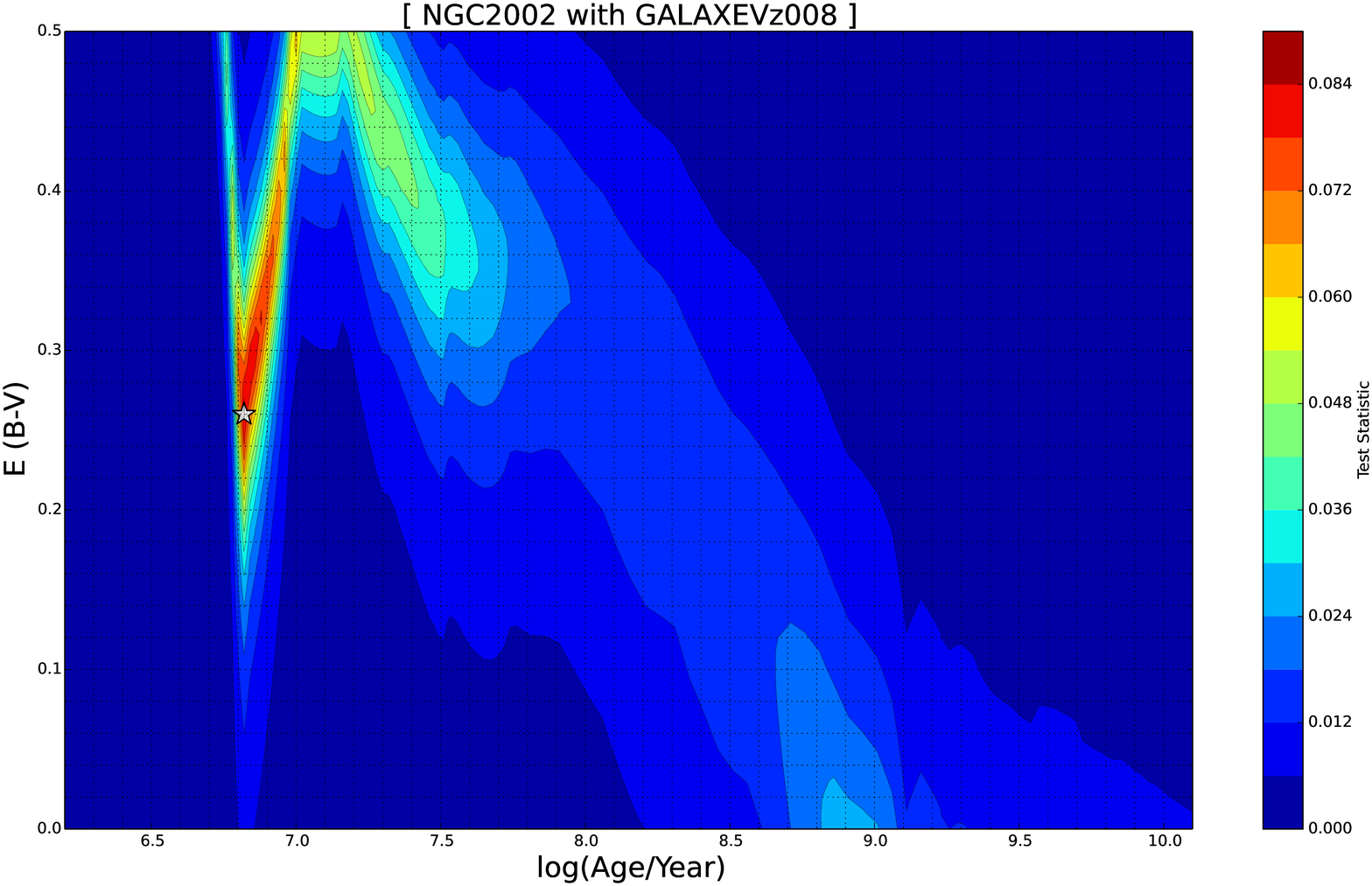}} &
\resizebox{75mm}{!}{\includegraphics{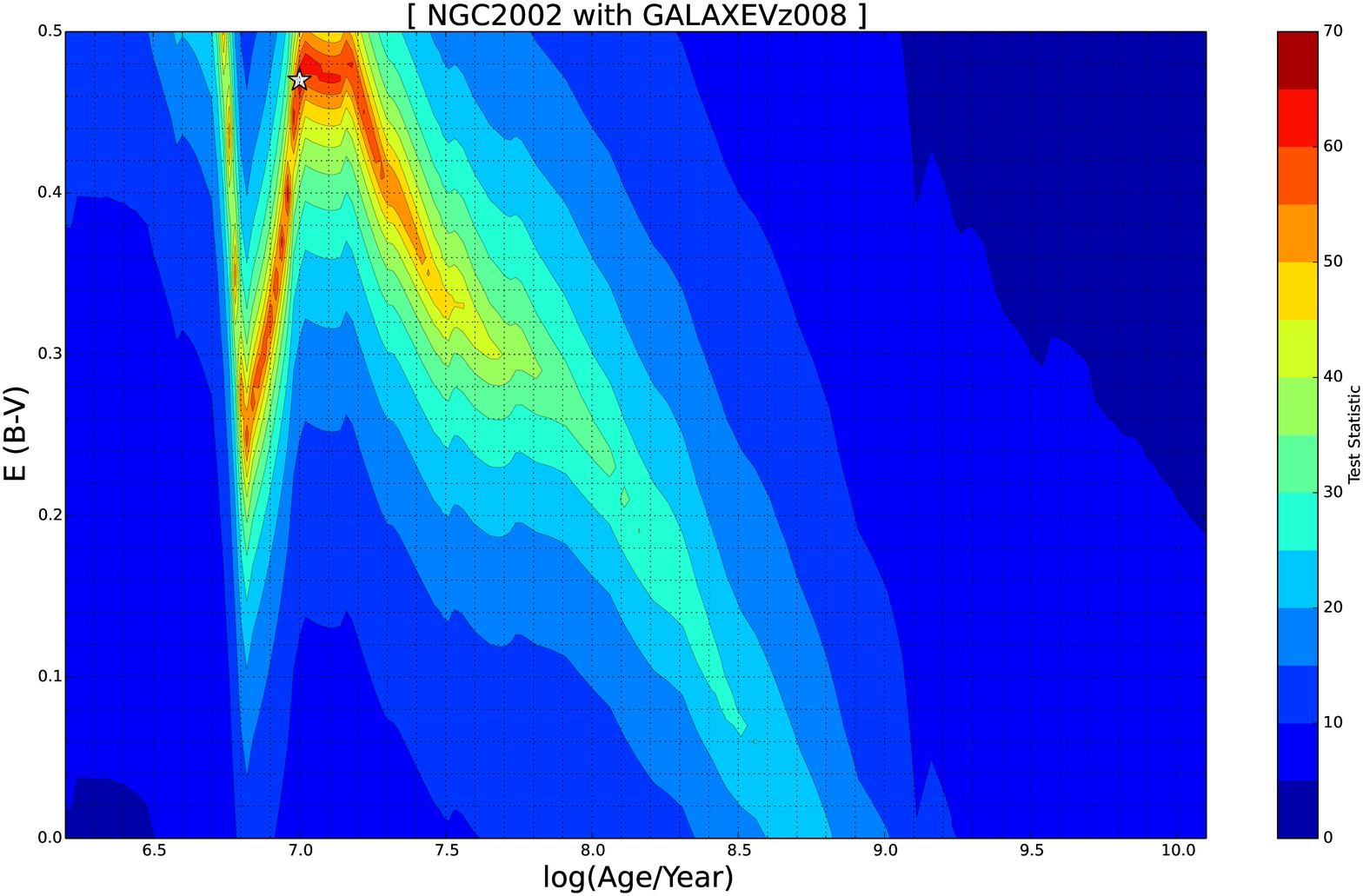}} \\

\end{tabular}
\caption{Results for NGC2002. The left column shows the results of the $\chi^{2}$ minimization method and the right column shows the results of the K-S test.}
\label{NGC2002spectra}
\end{figure}

NGC2002 has a CMD log (age/year) 7.20. Figure \ref{NGC2002spectra} shows the plots obtained by ASAD$_2$ for this cluster. GALAXEV cannot  produce a perfect spectral match because of the shape of the continuum of the observed spectrum. NGC2002 has an even greater flux offset between model and observed spectrum than NGC1994 because it has a worse S/N. The surface plots of GALAXEV show a range of possible solutions, the $\chi^{2}$ minimization solution is not unique (global minimum). Again we do not show the results from the MILES models as they cannot predict such young ages. 
\subsection {NGC2249}

\begin{figure}
\begin{tabular}{cc}
\resizebox{75mm}{!}{\includegraphics{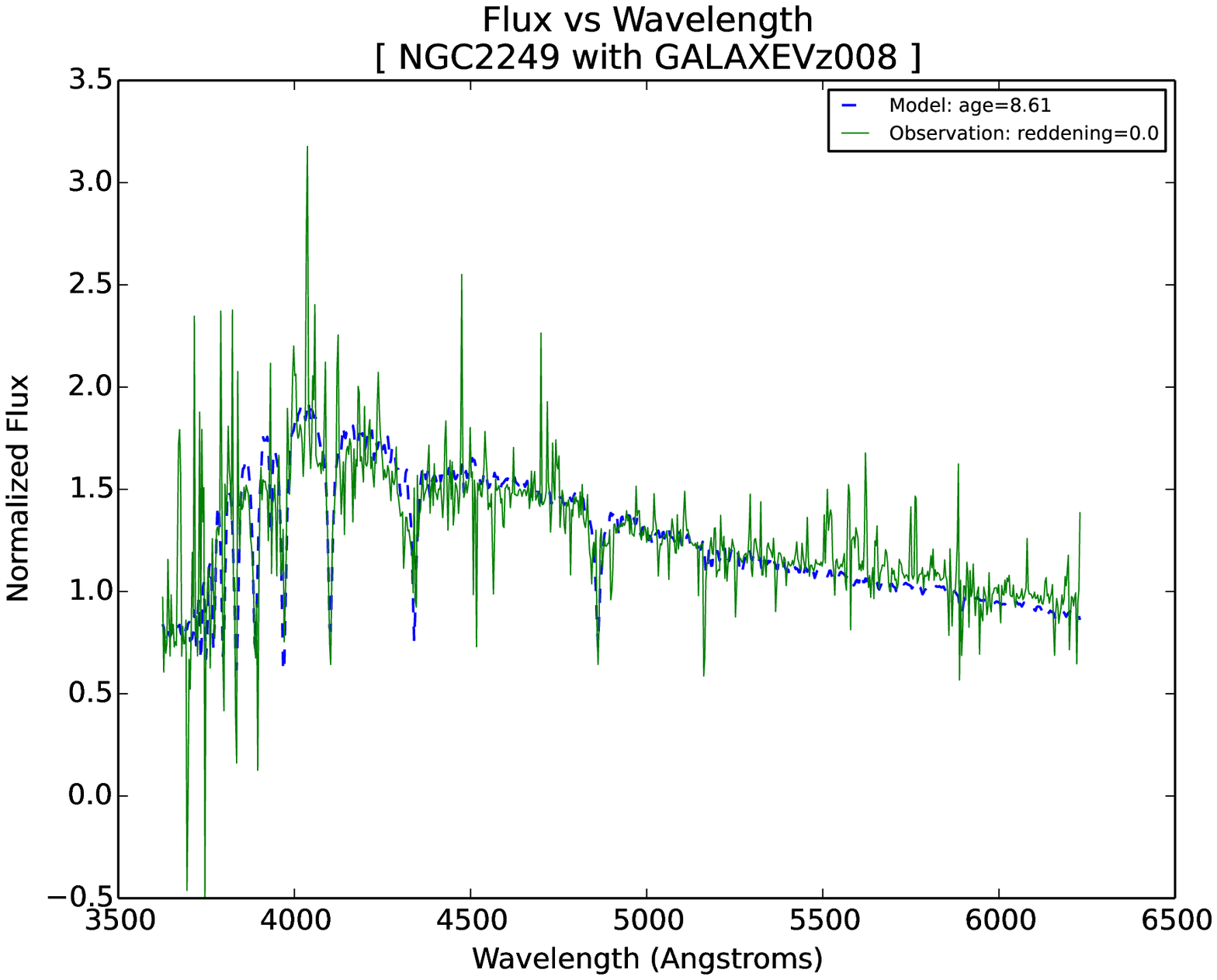}} &
\resizebox{75mm}{!}{\includegraphics{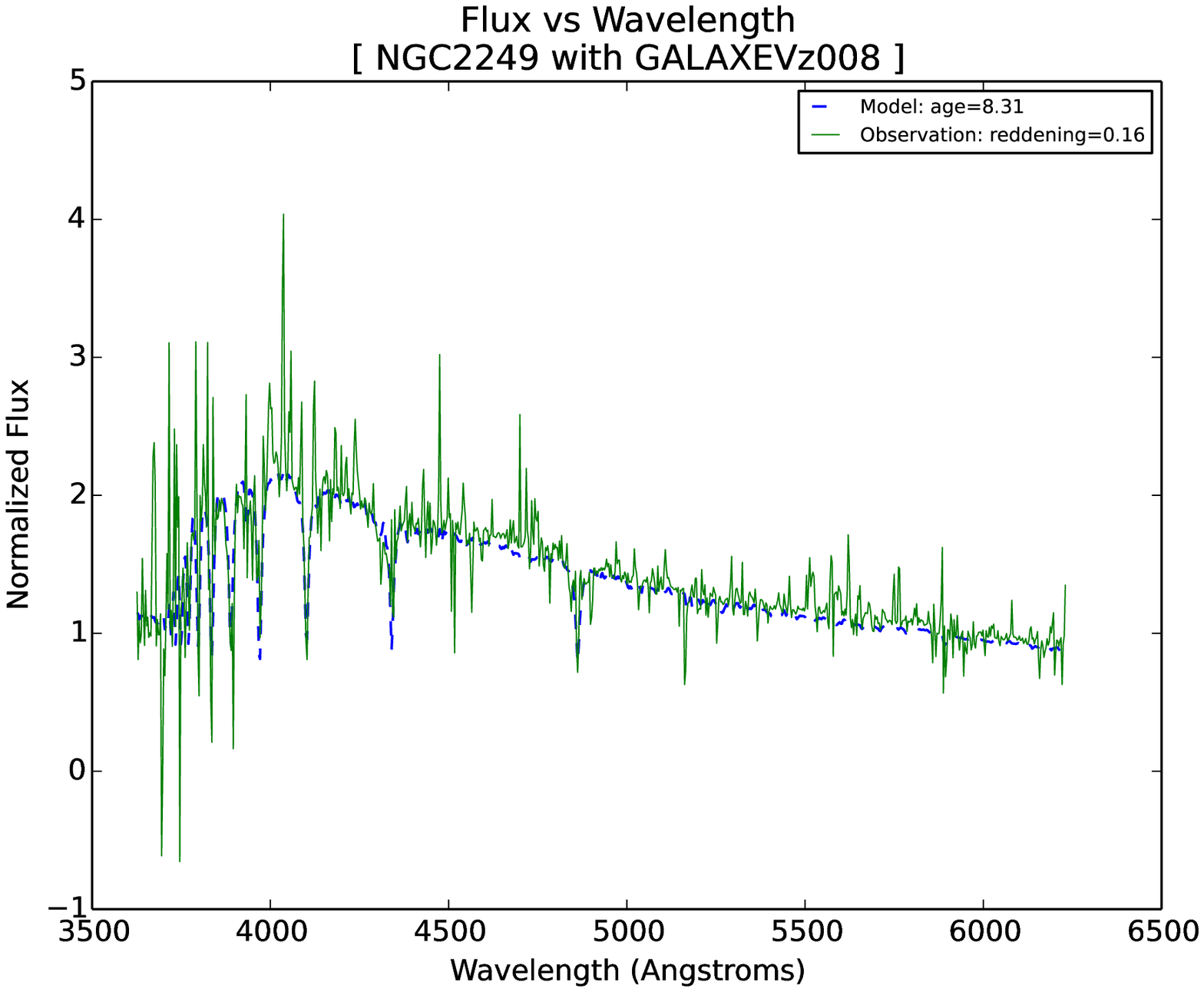}} \\
\resizebox{75mm}{!}{\includegraphics{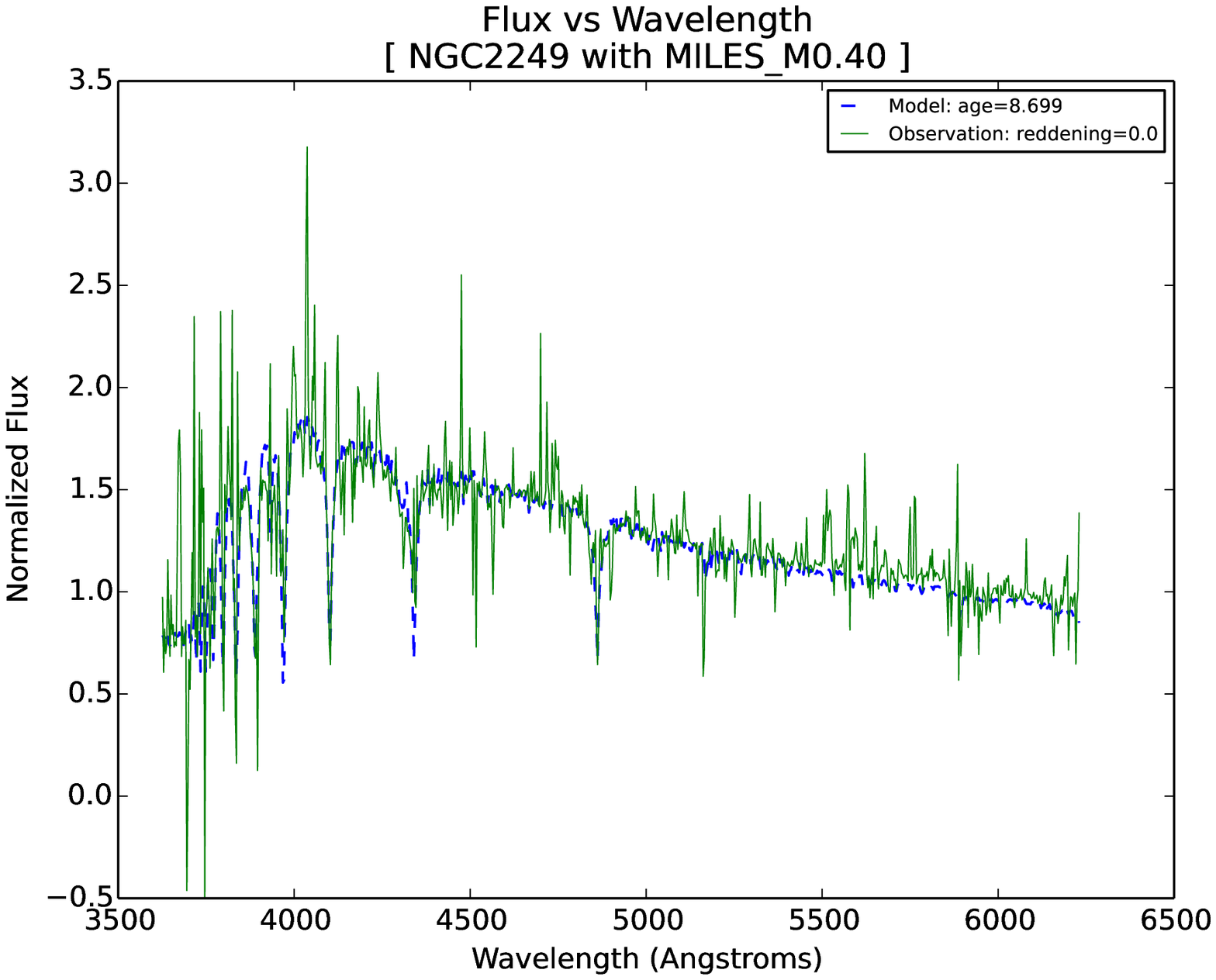}} &
\resizebox{75mm}{!}{\includegraphics{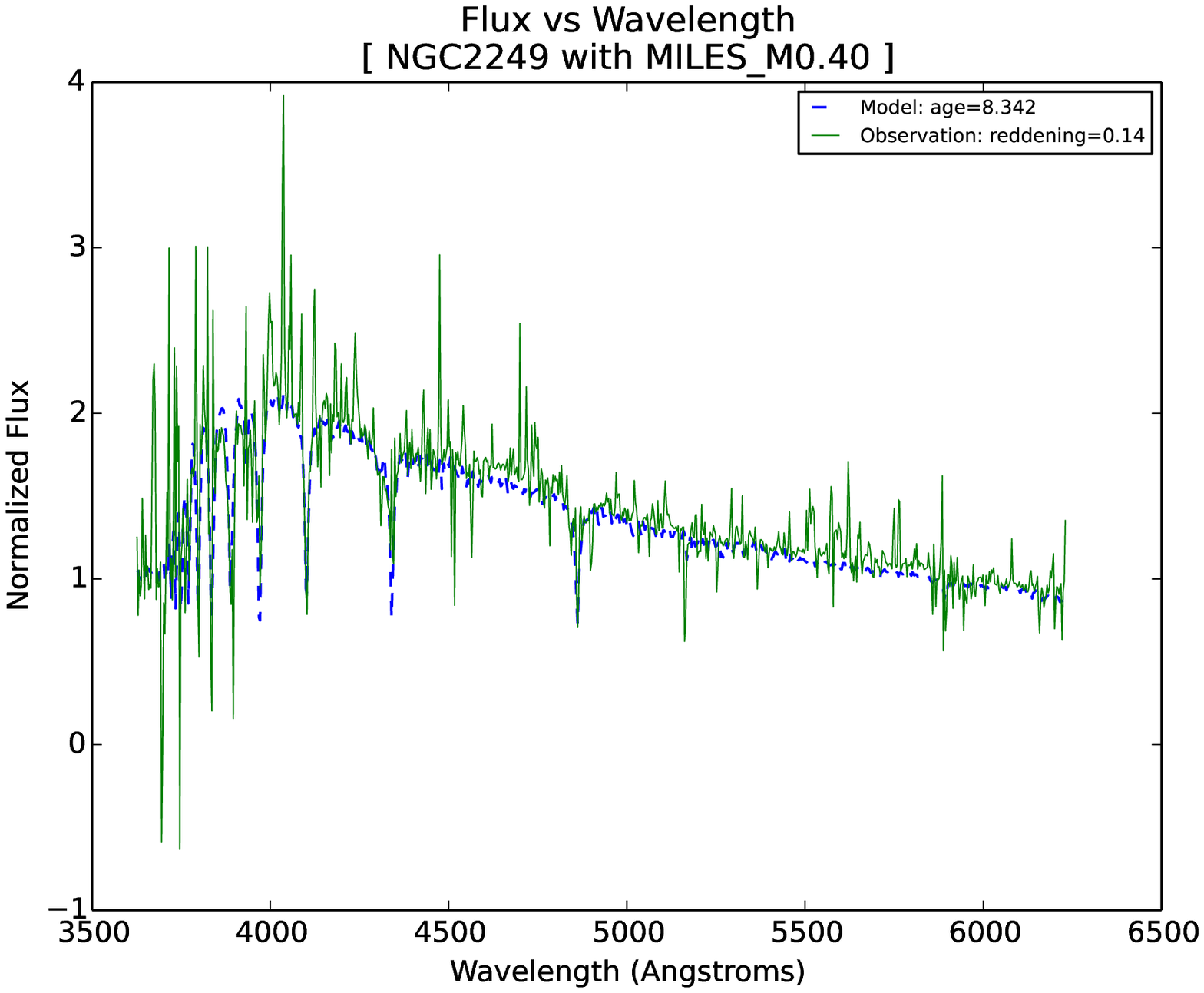}} \\
\resizebox{75mm}{!}{\includegraphics{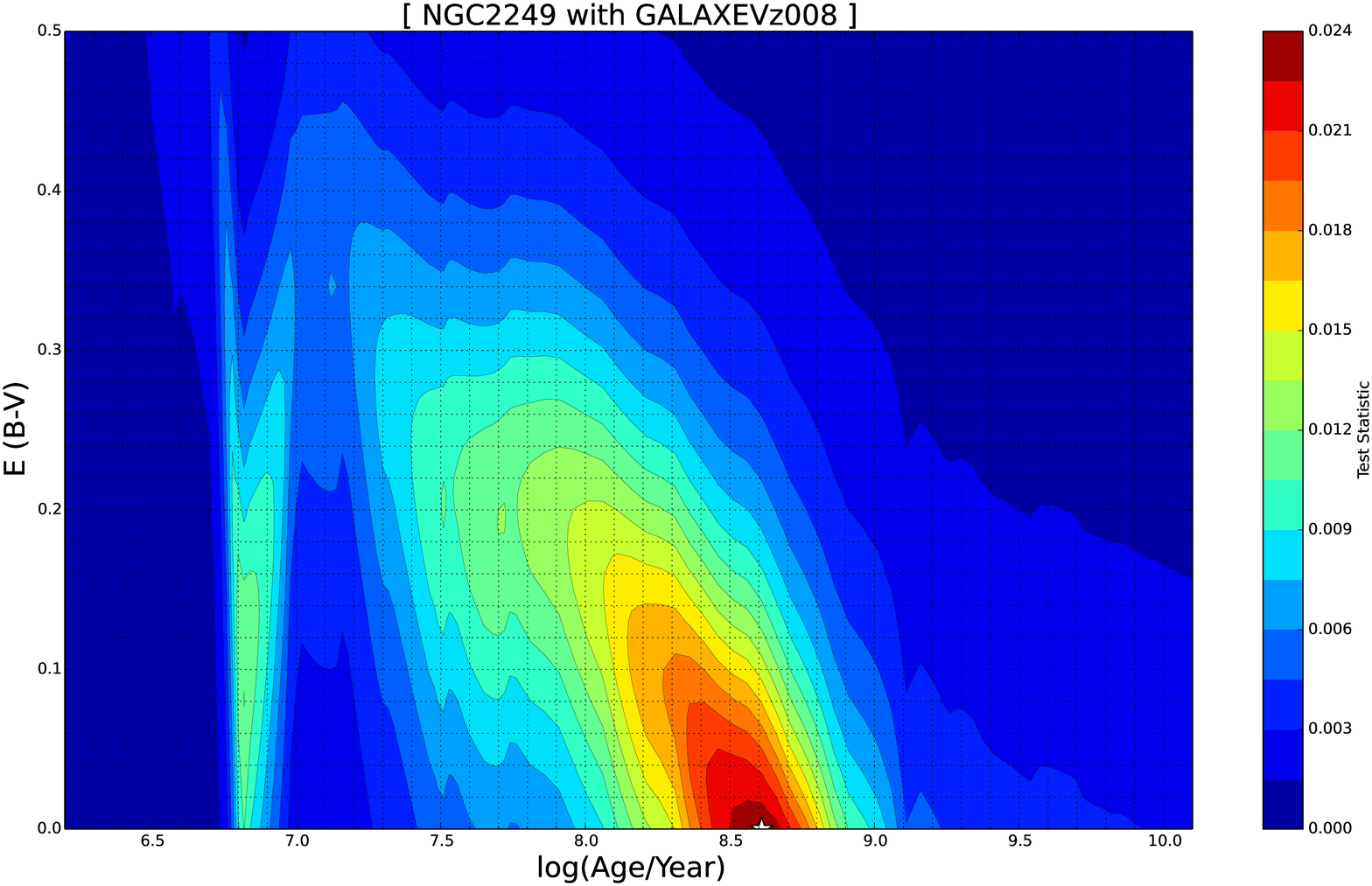}} &
\resizebox{75mm}{!}{\includegraphics{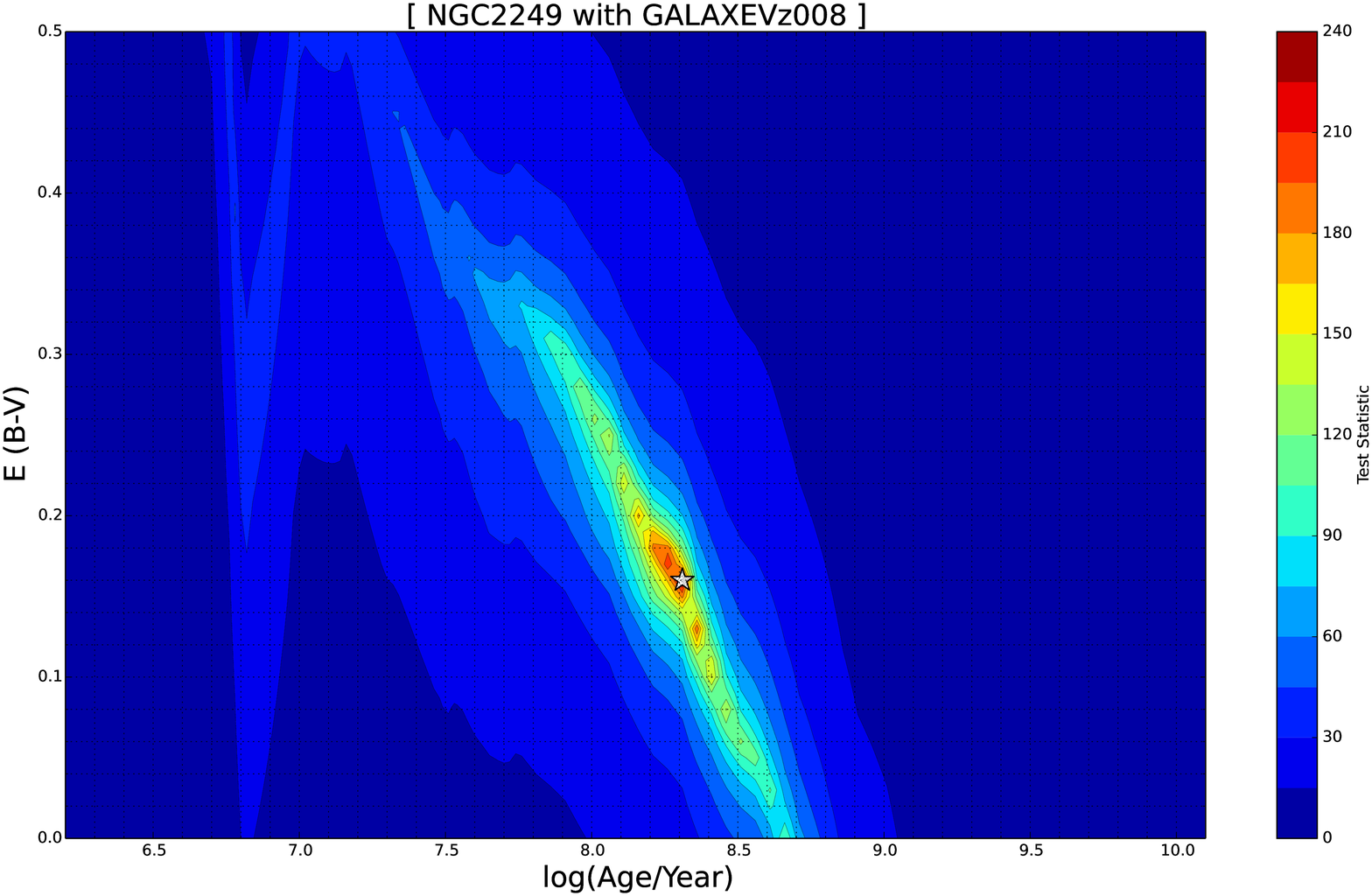}} \\
\resizebox{75mm}{!}{\includegraphics{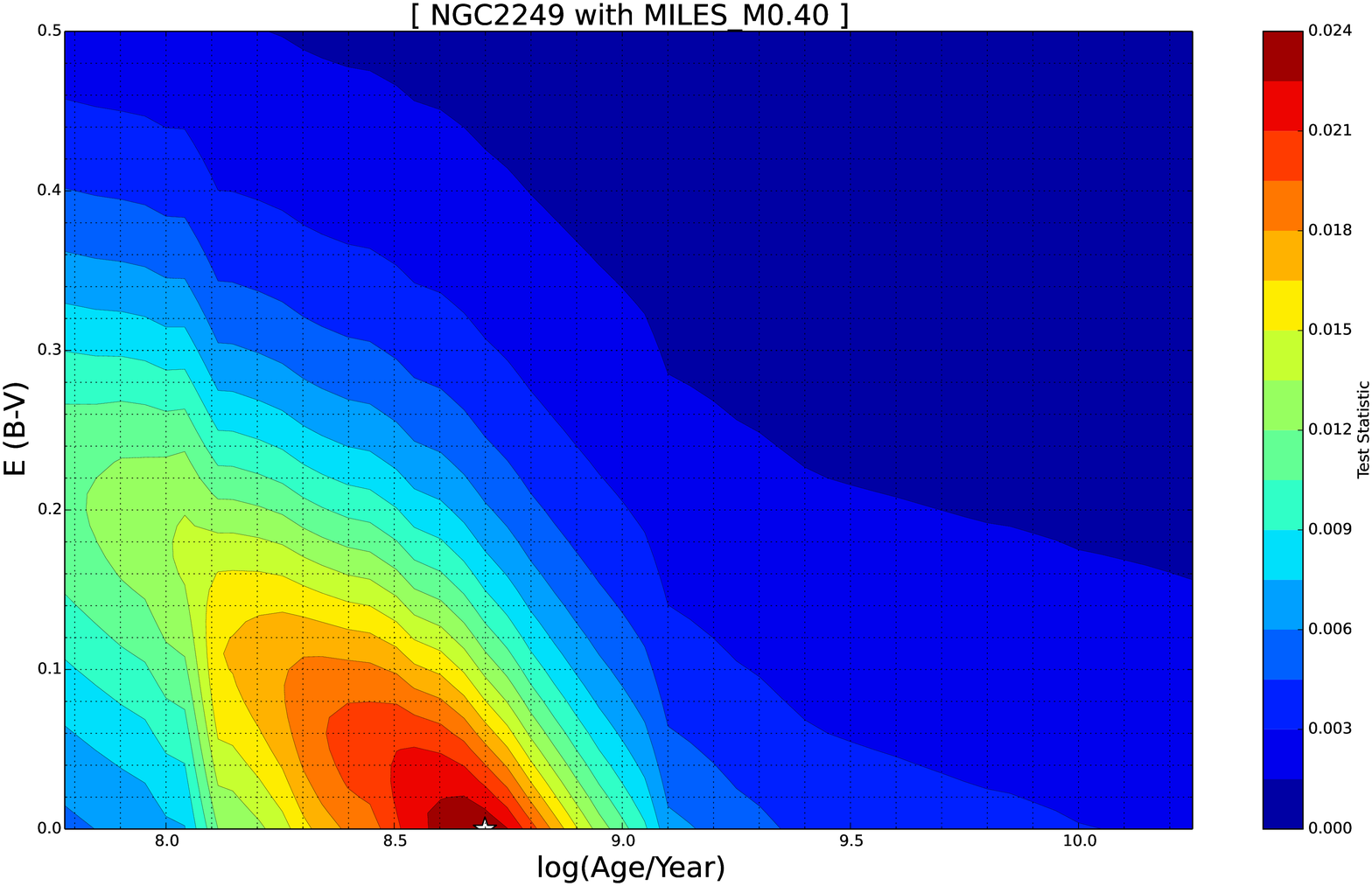}} &
\resizebox{75mm}{!}{\includegraphics{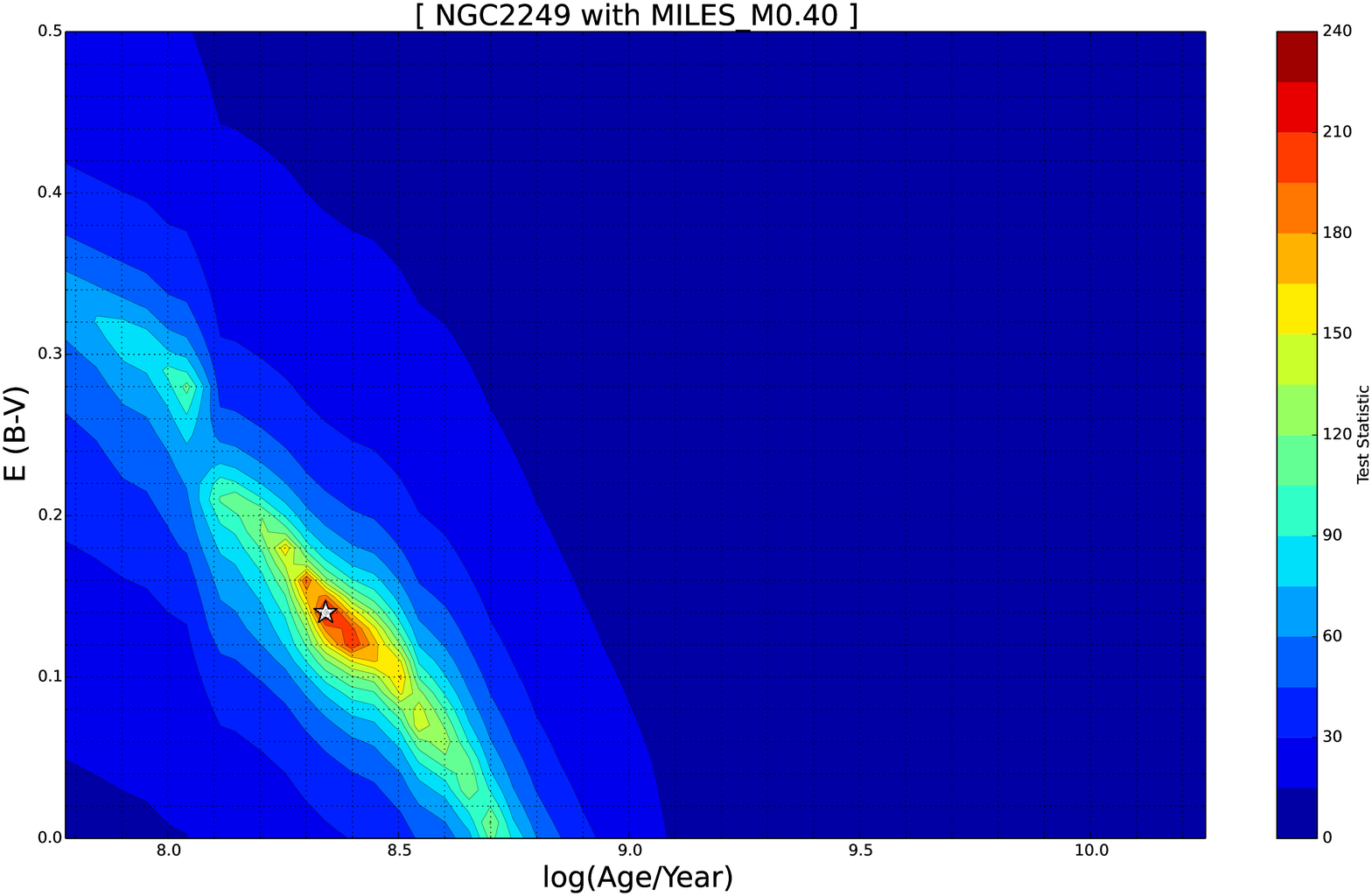}} \\
\end{tabular}
\caption{Results for NGC2249. The left column shows the results of the $\chi^{2}$ minimization method and the right column shows the results of the K-S test.}
\label{NGC2249spectra}
\end{figure}

NGC2249 has a CMD log (age/year) 8.82. Figure \ref{NGC2249spectra} shows the plots obtained by ASAD$_2$ for this cluster. The $\chi^{2}$ minimization method predicts ages closer to the CMD value. The K-S method predicts younger ages with greater reddening. 

\subsection {NGC2173}

\begin{figure}
\begin{tabular}{cc}
\resizebox{75mm}{!}{\includegraphics{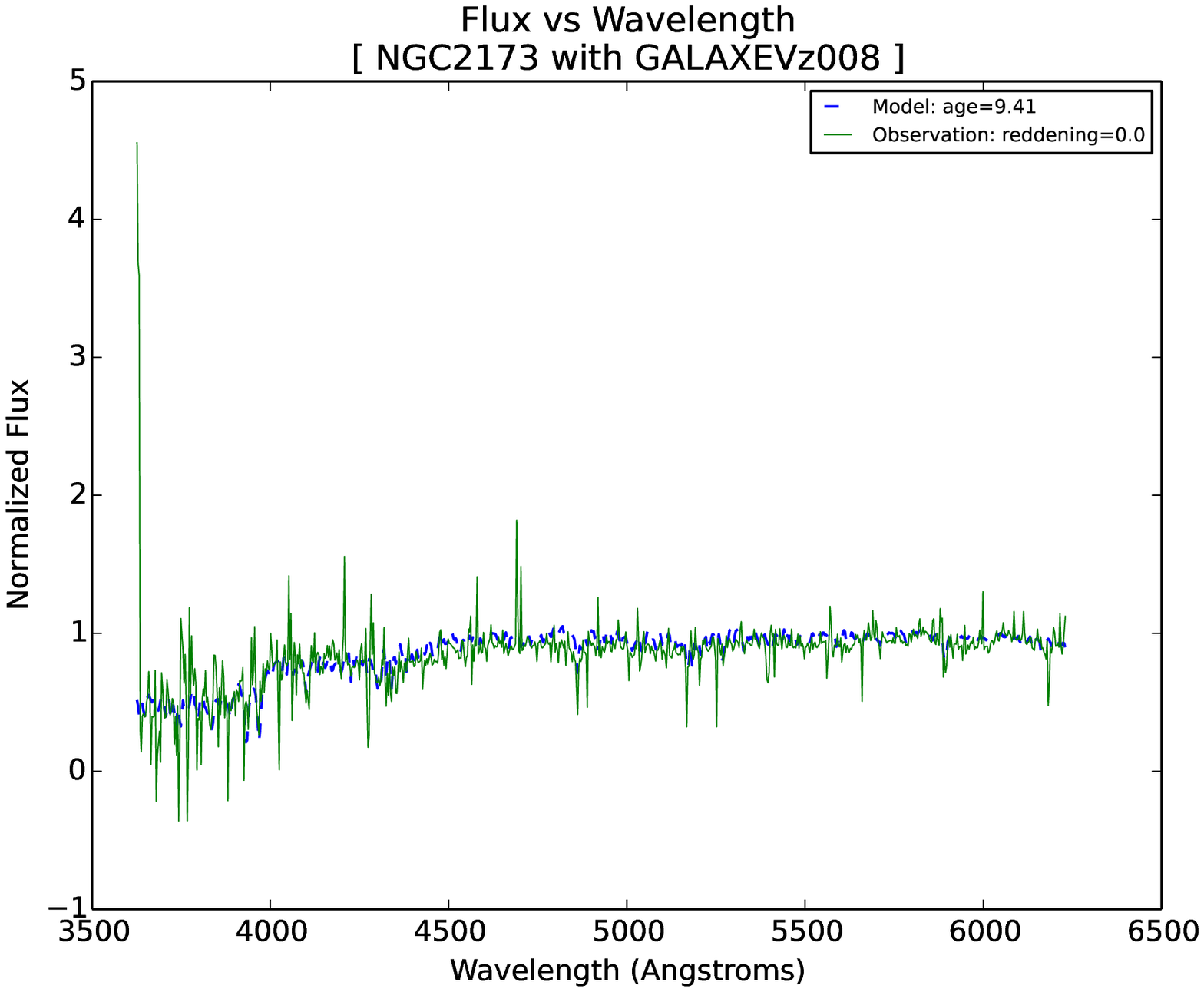}} &
\resizebox{75mm}{!}{\includegraphics{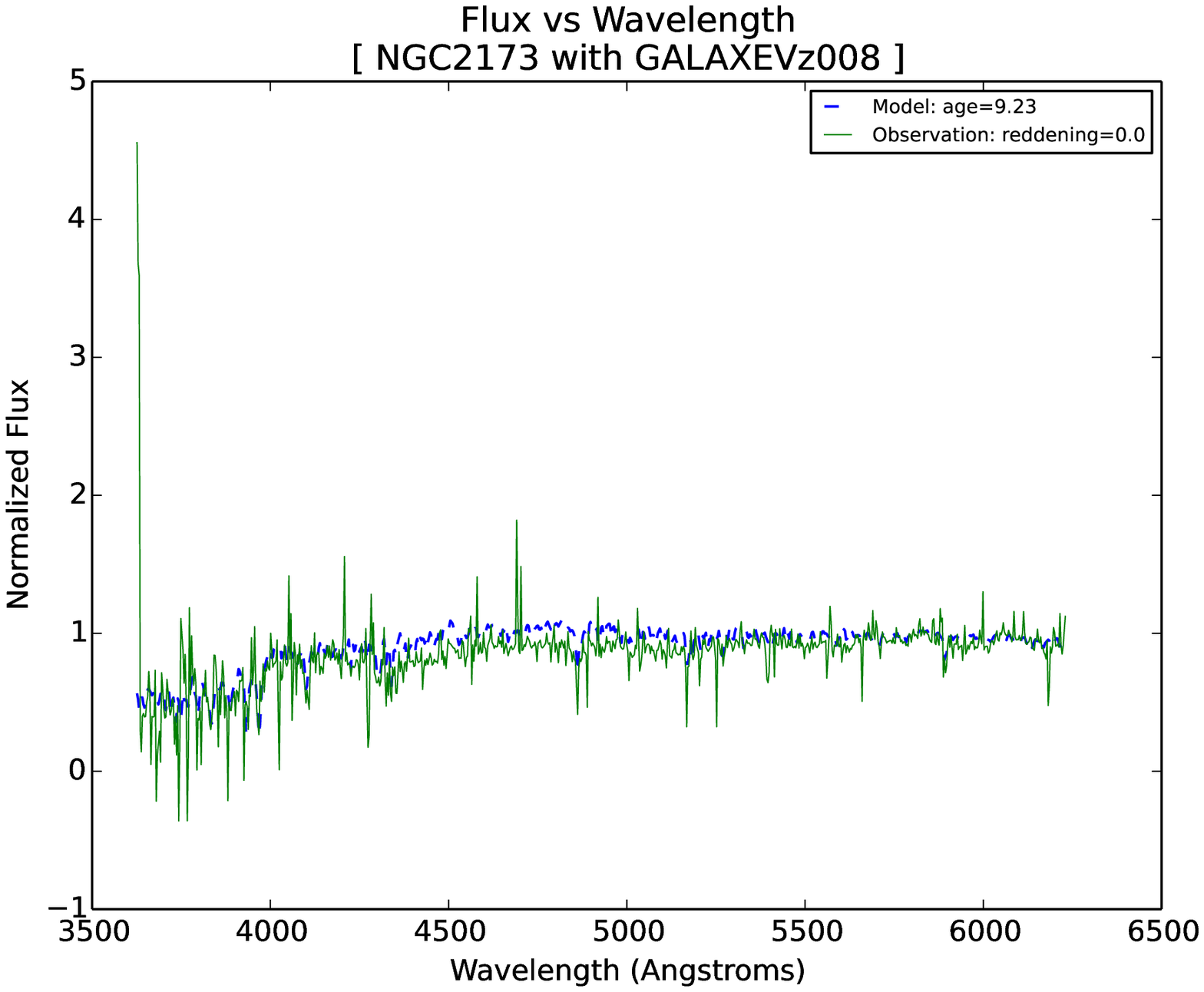}} \\
\resizebox{75mm}{!}{\includegraphics{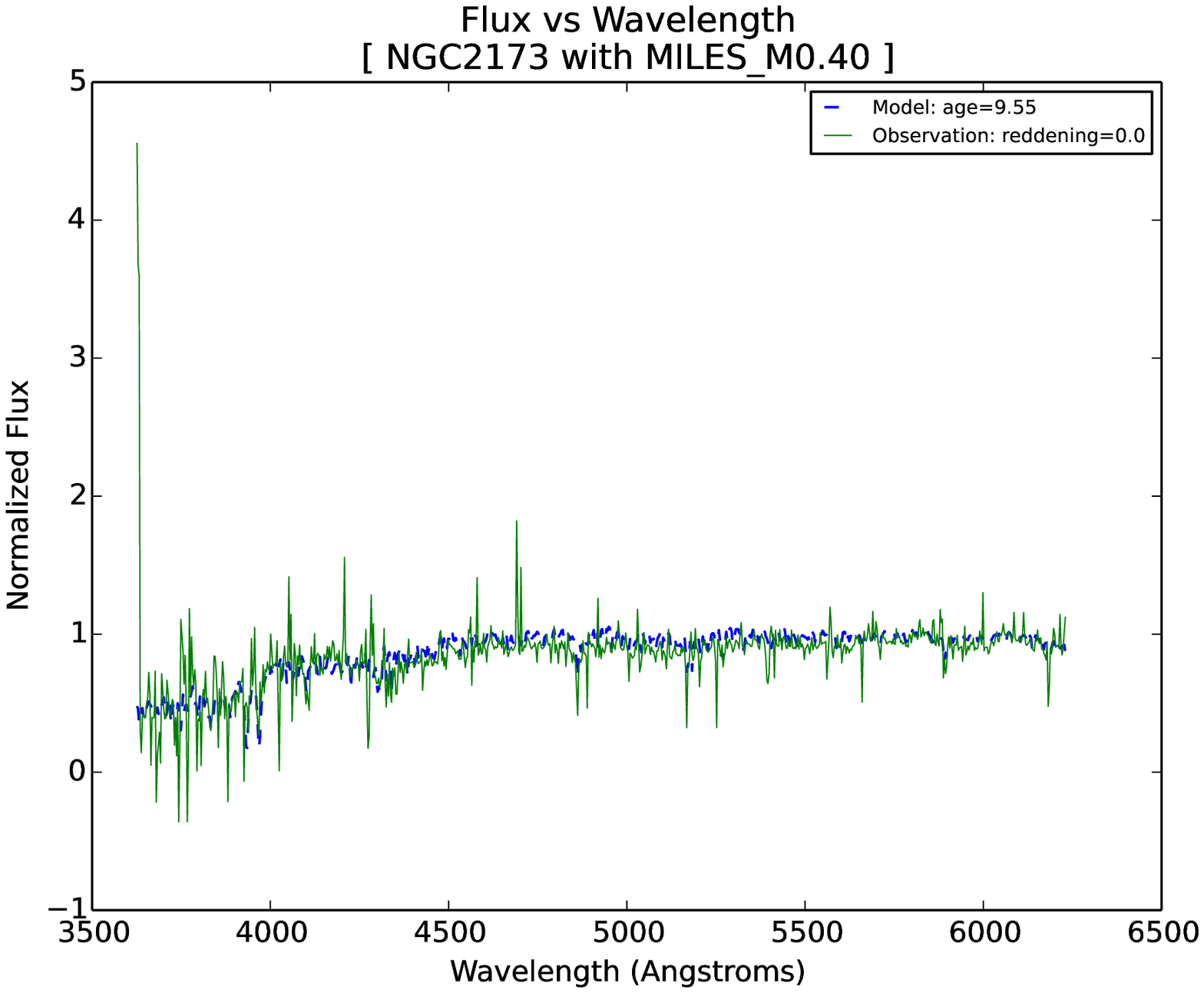}} &
\resizebox{75mm}{!}{\includegraphics{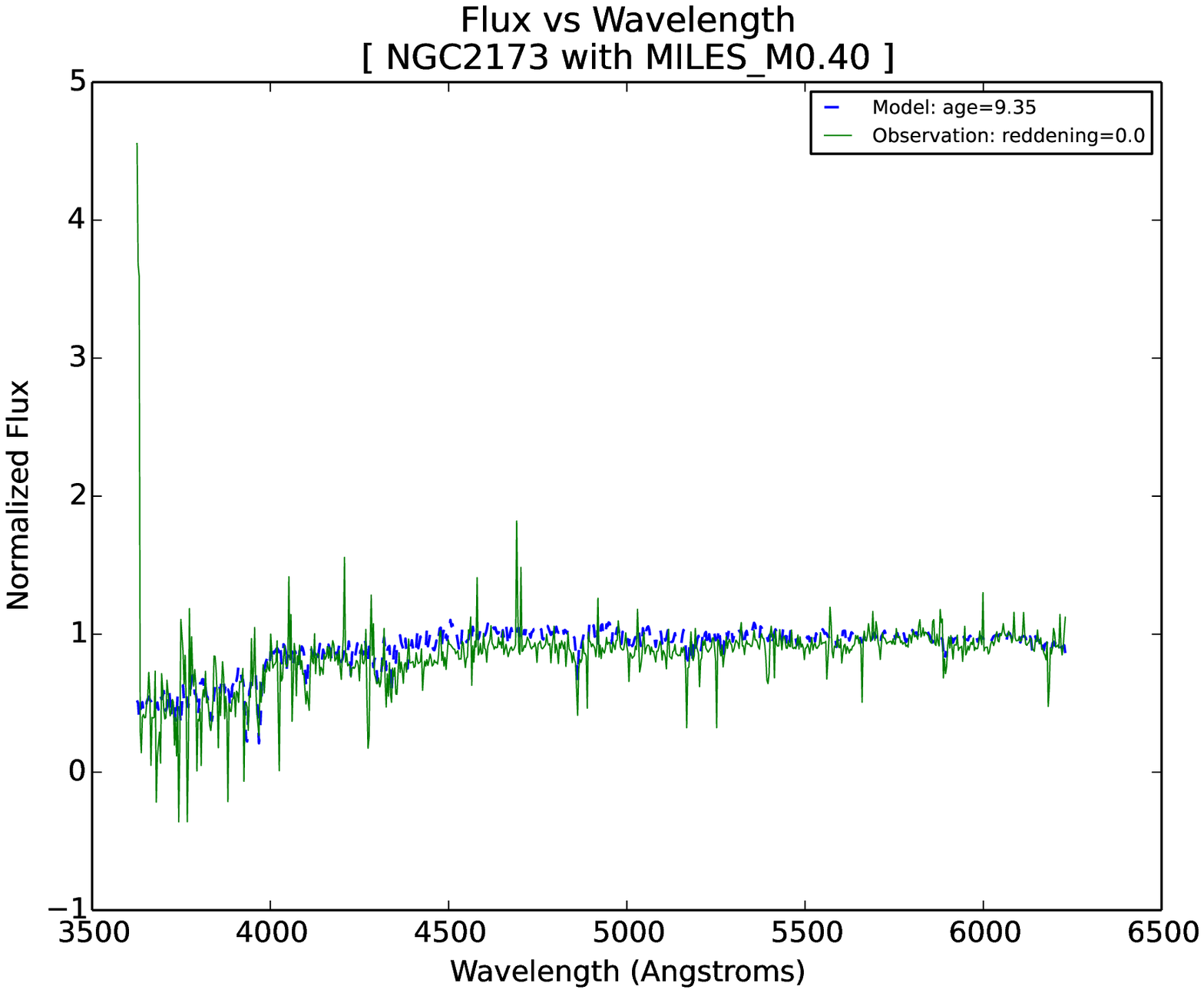}} \\
\resizebox{75mm}{!}{\includegraphics{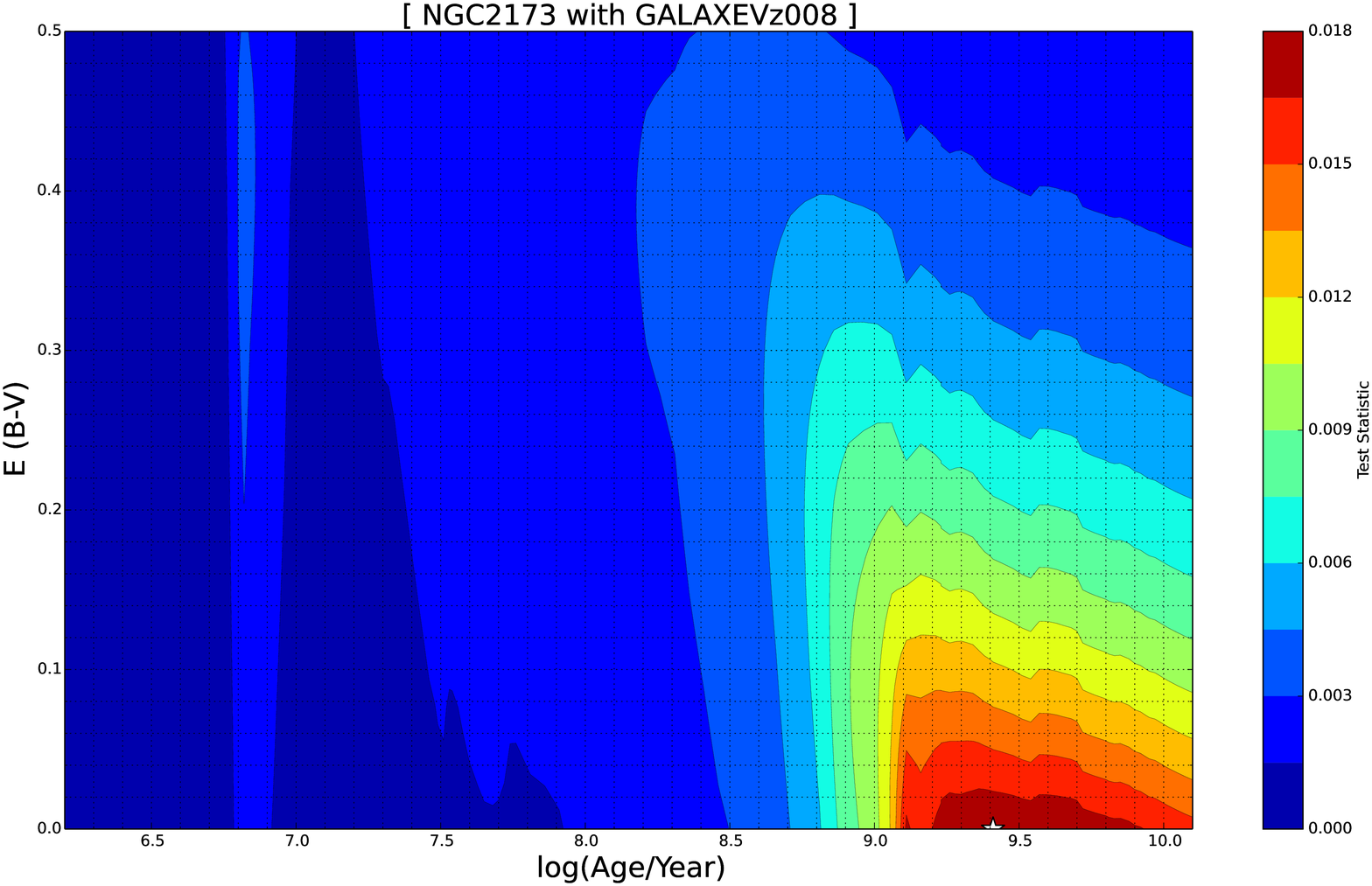}} &
\resizebox{75mm}{!}{\includegraphics{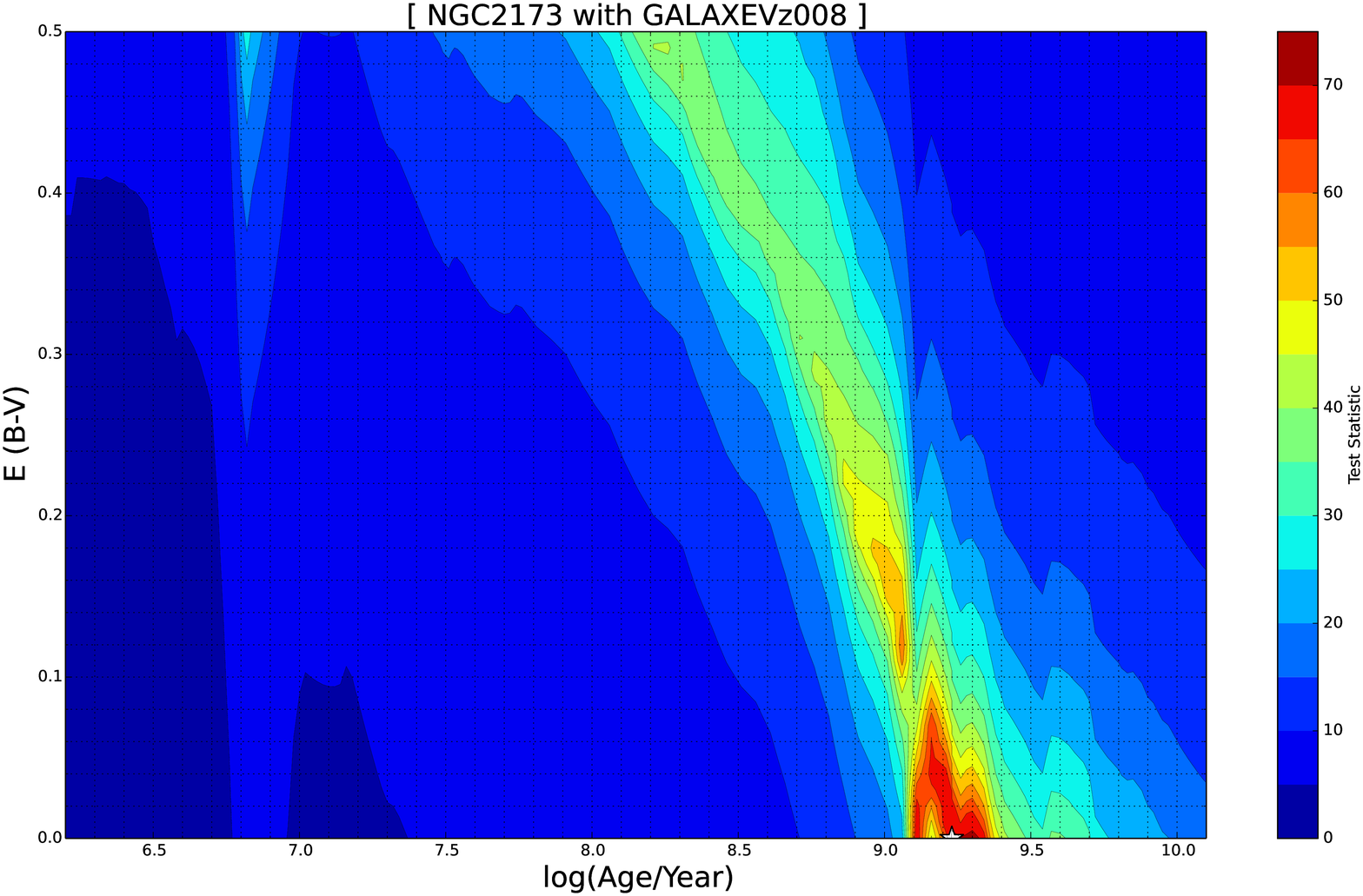}} \\
\resizebox{75mm}{!}{\includegraphics{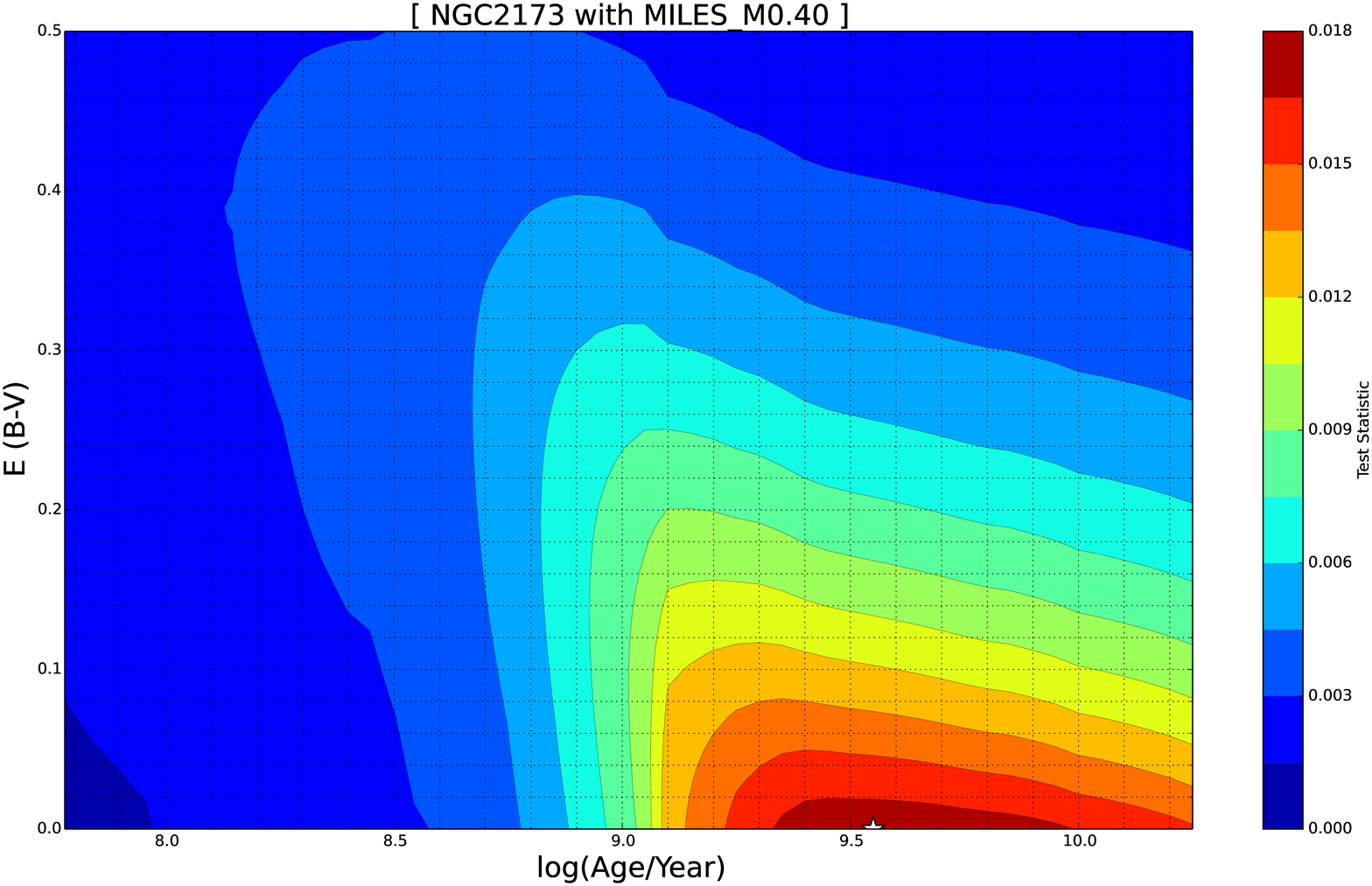}} &
\resizebox{75mm}{!}{\includegraphics{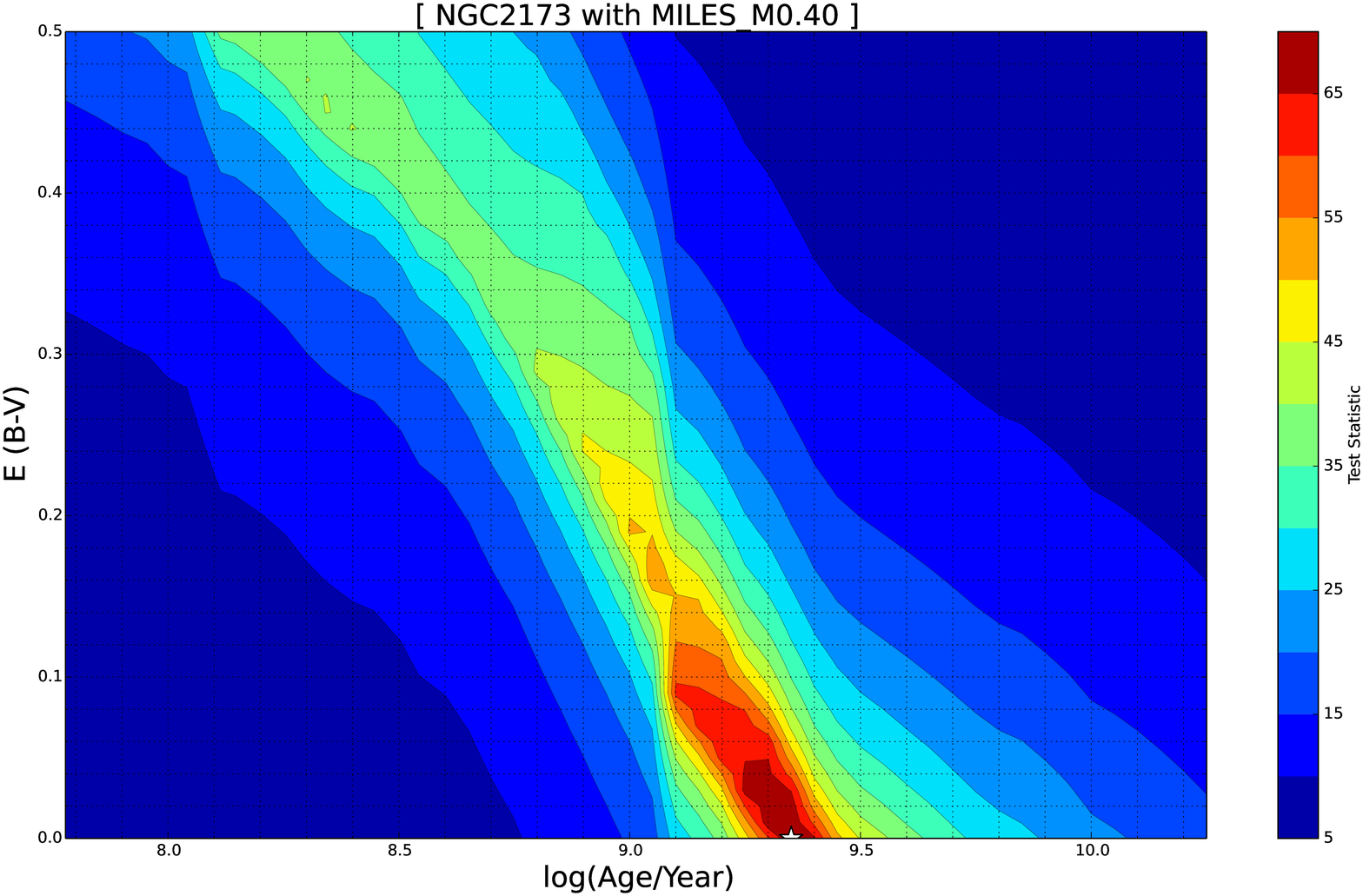}} \\
\end{tabular}
\caption{Results for NGC2173. The left column shows the results of the $\chi^{2}$ minimization method and the right column shows the results of the K-S test.}
\label{NGC2173spectra}
\end{figure}

NGC2173 has a CMD log (age/year) 9.32. Figure \ref{NGC2173spectra} shows the plots obtained by ASAD$_2$ for this cluster. The spectra show good match. The surface plots show that the solutions are unique (only one dark red region) but not precise (the dark red region has an extended area).

\section{Summary}

In this paper we presented ASAD$_2$ which is the updated version of ASAD \citet{Asad14}. We used it to conclude the following points:\\

1.Unlike the $\chi^{2}$ minimization method, the K-S method can predict reddening values higher than the values accepted for the LMC clusters. For one of the clusters in the sample, it also fails to break the age/reddening degeneracy.\\
2. Metallicity does not strongly affect the age determination for the full spectrum filling method regardless of the model used for log (age/year) $<$ 9. We are developing our method for older clusters, where the age/metallicity degeneracy is significant. For young cluster (log (Age/year) $<$ 7.4) the average age difference when comparing the age prediction of different metallicities is less than 0.1. For intermediate ages (7.4 $<$ log (Age/year) $<$ 8.8) the average difference is around 0.2 and for ages $>$ 8.8 the average difference is $>$ 0.35. In general, the average difference in log (Age/year) obtained by different metallicities with MILES is less than that seen for GALAXEV.  \\
3. There is a strong correlation between the ages predicted by Delgado model and GALAXEV. The difference in the predicted log (age/year) by the two models for 50\% of the clusters is less than 0.05. \\
4. There is a good agreement between the ages predicted by MILES and GALAXEV for ages greater than log (age/year) of 7.78, but because MILES does not have predictions for younger ages the estimated young ages don't match those of GALAXEV. \\
5. When comparing the results obtained with MILES models using a fixed resolution higher than that of the data, to those obtained when matching the MILES resolution to the data resolution, we notice that the results are almost identical. \\
6. When comparing the age prediction difference of $|$(KS method - $\chi^{2}$ minimization method)$|$ versus CMD log (age/year), the least difference is seen for intermediate ages 7 $<$ log (age/year) $<$ 8.5.\\ 
7. When comparing the age prediction difference of $|$(KS method - $\chi^{2}$ minimization method)$|$ versus S/N,  the difference vary for S/N $<$ 60 and it is minimum for S/N $>$ 60.\\
8. When comparing the age prediction difference of $|$(KS method - $\chi^{2}$ minimization method)$|$ versus resolution (FWHM), for GALAXEV the difference decreases as the resolution increase. For MILES the difference is the same for the two resolutions 14$\rm\AA$ (FWHM) and 17$\rm\AA$ (FWHM).\\
9. The sharp spectrum transition originated at supergiant and AGB ages are limiting our ability to provide values in agreement with the CMD estimates and as a result the reddening determination is not accurate.

We thank Adnan Shahpurwala who carefully ran the Windows version of ASAD$_2$ and compared the results. His availability to always help is highly appreciated. We thank the anonymous referee for providing constructive comments for improving the content of this paper. We also thank Dr. Santos and Dr. Palma for allowing us to use their integrated spectra for 7 clusters. This material is based upon work supported in part by the FRG14-2-05 Grant P.I., R.\ Asa'd from American University of Sharjah. We also acknowledge support from grant AYA2013-48226-C3-1-P from the Spanish Ministry of Economy and Competitiveness (MINECO).

\clearpage

\clearpage

\clearpage

\clearpage

\clearpage

\clearpage

\clearpage

\clearpage

\bibliographystyle{apj}     
\bibliography{NewestFeb2015}

\end{document}